\documentclass[journal=jacsat,manuscript=article]{achemso}
\usepackage{lineno}
\usepackage{amssymb}
\usepackage[version=3]{mhchem} % Formula subscripts using \ce{}

\author{Matteo Castellani}
\author{Owen Medeiros}
\author{Alessandro Buzzi}
\author{Reed A. Foster}
\author{Marco Colangelo}
\author{Karl K. Berggren}

\affiliation[Massachusetts Institute of Technology]
{Dept. of Electrical Engineering and Computer Science, Massachusetts Institute of Technology, Cambridge, MA, USA}

\email{mcaste@mit.edu}

\title[An \textsf{achemso} demo]
  {A superconducting full-wave bridge rectifier}

\abbreviations{IR,NMR,UV}

\begin{document}

% maximum 150 words
\begin{abstract}
 
Superconducting thin-film electronics are attractive for their low power consumption, fast operating speeds, and ease of interface with cryogenic systems such as single-photon detector arrays, and quantum computing devices.
However, the lack of a reliable superconducting two-terminal asymmetric device, analogous to a semiconducting diode, limits the development of power-handling circuits, fundamental for scaling up these technologies.
Existing efforts to date have been limited to single-diode proofs of principle and lacked integration of multiple controllable and reproducible devices to form complex circuits.
Here, we demonstrate a robust superconducting diode with tunable polarity using the asymmetric vortex surface barrier in niobium nitride micro-bridges, achieving a 43\,\%  peak rectification efficiency, and showing half-wave rectification up to 120\,MHz.
We then realize and integrate several such diodes into a bridge rectifier circuit on a single microchip that performs continuous full-wave rectification up to 3\,MHz and AC-to-DC conversion of a 50\,MHz signal in periodic bursts with an estimated peak power efficiency of 50\,\%.
\end{abstract}

\newpage

Superconducting nanowire single-photon detectors (SNSPD) arrays are expected to be of key importance to the future development of photonic quantum computing \cite{alexander_manufacturable_2025}, quantum communication \cite{grunenfelder_fast_2023}, optoelectronic spiking neural networks \cite{khan_superconducting_2022}, and high-energy physics \cite{lee_beam_2024}.
Scaling up these systems requires the integration of cryogenic co-processors for readout and signal processing \cite{oripov_superconducting_2023, castellani_nanocryotron_2024, huang_monolithic_2024, miyajima_high-time-resolved_2018, yabuno_scalable_2020, viskova2022cryo}, and interconnection between many modular units as in superconducting quantum computing platforms \cite{acharya_multiplexed_2023, niu_low-loss_2023}.
Therefore, it is crucial to build architectures capable of delivering and handling significant power at cryogenic stages, while dynamically tuning bias levels, minimizing cross-talk and electrical noise, and maintaining thermal isolation. 
Any solution to this problem will have to leverage the intrinsic energy efficiency of thin-film superconductors, be compatible with standard materials used for superconducting detectors and electronics, and interface with conventional circuitry.

%diodes are needed for power
In modern integrated circuits, diodes are fundamental components for power control and distribution. 
As such, superconducting diodes will be needed for efficient and compact bias-handling systems needed in cryogenic electronics and detectors. 
The main functionality enabled by superconducting diodes would be AC-to-DC conversion at low temperatures (e.g. using full-wave bridge rectifiers) to provide stable and tunable DC bias currents through RF signals. 
Such an operation might be useful to deliver power from room temperature using a limited number of RF lines or to interconnect different sub-systems of the same cryogenic architecture.
Other typical functionalities of semiconducting diodes could be realized with superconductors, such as frequency mixing, signal modulation, voltage-controlled oscillation, and isolation; benefiting the development of several cryogenic platforms. 

The superconducting diode effect (SDE) manifests in asymmetric critical currents for opposite current directions in a superconducting device.
This phenomenon has been recently demonstrated using multiple technologies \cite{nadeem_superconducting_2023} that primarily rely on the combined effect of a magnetic field and spin-orbit coupling in heterostructured materials \cite{ando_observation_2020, bauriedl_supercurrent_2022} and Josephson junctions \cite{baumgartner_supercurrent_2022, pal_josephson_2022}, or on the influence of a magnetic field on multi-junction devices \cite{paolucci_gate-_2023, golod_demonstration_2022}. 
Moreover, field-free superconducting diodes have also been realized \cite{wu_field-free_2022, narita_field-free_2022}. 

% motivation: realistic circuits
Despite impressive technical progress in the formation of individual devices, and the recent demonstration of half-wave rectification at 100\,kHz,\cite{chahid_high-frequency_2022} diode effects have not been explored for near-term use in complex and useful circuits that require more than a single diode. Past diode demonstrations had limitations such as: low rectification efficiency ($\eta$), the need for Tesla-scale applied magnetic field, or the use of complex geometries and heterostructures requiring materials processing that is incompatible with conventional superconducting circuits.
Thus, the assumption that cryogenic circuits could one day usefully take advantage of these SDEs remains to be proven. 

% vortex diode in nbn is a good option for these two reasons, 
% story of the effect (short)
Among the variety of proposed superconducting diode effects, the vortex diode effect, based on the asymmetric vortex surface barrier in micro-bridges,  has a great potential for realizing diode-based circuits at a large scale because: it requires the patterning of only a single layer; it operates with milliTesla-scale fields; it can generate a high-resistance state (useful to interface with traditional electronics) and; it can be implemented in essentially any superconductor. 
In particular, implementing vortex diodes in niobium nitride thin films opens the door to a wide range of applications considering that such material has been widely used for high-performance SNSPDs \cite{korzh_demonstration_2020}, nanocryotron digital circuits that can be monolithically integrated with SNSPDs \cite{mccaughan_using_2016, butters_scalable_2021, buzzi_nanocryotron_2023, foster_superconducting_2023, castellani_nanocryotron_2024, huang_monolithic_2024}, microwave devices \cite{colangelo_compact_2021, wagner_demonstration_2019}, spiking neural networks \cite{toomey_superconducting_2020, lombo_superconducting_2022}, and comparators that interface Josepshon junctions with conventional electronics \cite{zhao_nanocryotron_2017}.

%literature eview of vortex diodes, problems to solve to do a NbN vortex diode
The vortex diode effect was initially proposed by Vodolazov et al.\cite{vodolazov_superconducting_2005}, and then observed by Cerbu et al. \cite{cerbu_vortex_2013} in aluminum wires, by Semenov et al. in NbN SNSPDs with $90^{\circ}$ turns \cite{semenov_asymmetry_2015}, and by Hou et al. \cite{hou_ubiquitous_2024} with controlled edge defects on straight vanadium micro-bridges under an applied out-of-plane field of a few milliTesla.  In this device, the efficiency was also boosted from $50\%$ to $65\%$ by introducing an adjacent thin layer of ferromagnetic material as done by Gutfreund et al. \cite{gutfreund_direct_2023},  to eliminate the need for geometrical defects and external fields.
The vortex diode effect has been recently demonstrated also using NbTiN nanowires \cite{zhang_superconducting_nodate} ($\eta = 24$\,\%), and NbN micro-bridges ($\eta = 30$\,\%) \cite{suri_non-reciprocity_2022}; however, the NbN-based devices relied on random edge defects introduced during the fabrication process, thus limiting the device reproducibility. 
Overcoming this limitation would help realize a reliable and reproducible NbN diode useful to design AC-to-DC converters in power distribution architectures.

%what we did practically
In this work, we demonstrate the potential of using vortex diodes for designing power-handling systems by combining them into a non-trivial circuit: a full-wave bridge rectifier composed of four integrated operational diodes. 
With this circuit, we demonstrated sustained full-wave rectification of sinusoidal signals at 3\,MHz; rectification of a continuous pulsing waveform at 20\,MHz; and AC-to-DC conversion at 50\,MHz for limited time intervals.  
We implemented the vortex diodes using micro-bridges with controlled triangular edge defects, following Hou et al. \cite{hou_ubiquitous_2024} but using NbN as the superconducting material. 
We characterized both stand-alone diodes and bridge rectifiers by studying how the rectification efficiency and input current margins change as a function of the field, frequency, and shape of the signal.
Additionally, we demonstrated half-wave rectification up to 120\,MHz and we investigated possible thermal effects limiting speed and margins of the devices. 
We also created an LTspice model of the diode to estimate the power efficiency of the AC-to-DC conversion and facilitate the design of larger diode-based circuits. 
%bias network
To show the prospect of using such devices in circuits at a larger scale, we designed and simulated an architecture, based on the demonstrated AC-to-DC converter, to bias a network of 10 SNSPDs with dynamically tunable DC currents, by frequency multiplexing the bias levels on a single RF line as described in the supplementary material.

\section{Bridge rectifier with vortex diodes}
   
    \begin{figure*}[htbp!]
      
        \includegraphics[width=14cm]{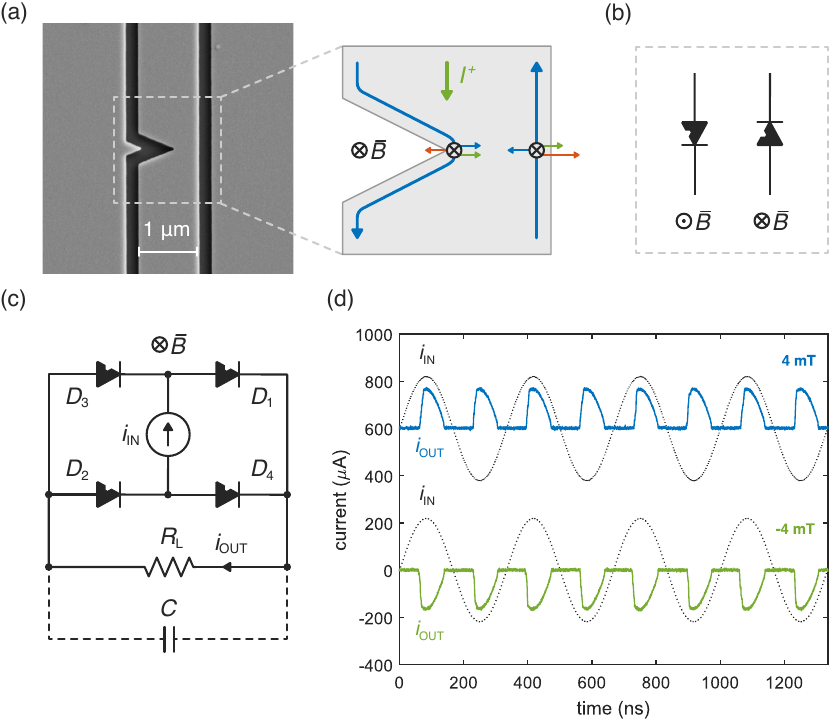}
        \caption{Superconducting diode and bridge rectifier. (a) Scanning electron micrograph of a superconducting diode, fabricated on a 14\,nm thick film of niobium nitride. The constriction is 400\,nm wide and the notch has an angle of $45^{\circ}$. The gray regions are niobium nitride, and the dark borders of the traces show the underlying SiO$_2$ substrate. The ground plane surrounds all the traces. The right inset depicts the working principle of a vortex diode when an entering out-of-plane magnetic field is applied, and a current $I^{+}$ passes through the device. Vortex dynamics are governed by three forces: force due to circulating current formed by the Meissner effect (horizontal blue arrows); Lorentz force due to $I^{+}$ (horizontal green arrows) and; force due to the surface barriers (red arrows). The net force pushes vortices from left to right. (b) Circuit symbols for superconducting diodes with positive or negative applied field. The field is entering the plane and the notch is on the left edge of the device. Inverting the field orientation changes the polarity of the diode. (c) Circuit schematic of the superconducting bridge rectifier driving a resistive load $R_\text{L}$. A smoothing capacitor $C$ (dashed line) can be added in parallel to $R_\text{L}$ to obtain a DC output current for AC-to-DC conversion. (d) Experimental full-wave current rectification of a 3\,MHz sinusoidal signal $i_\text{IN}$ (black dashed traces), with $R_\text{L} = 50$\,$\Omega$ and without smoothing capacitor, and for both positive (blue trace) and negative (green trace) applied field. Traces are vertically shifted for clarity. Each trace is the average of 10 different traces acquired in sequence, to increase the signal-to-noise-ratio (one of the 10 traces is shown in the supplementary material).}\label{fig: 1} 
    \end{figure*}

    %description of the diode and operation principle
    Our vortex diodes were 1\,µm wide wires patterned on NbN films with three different thicknesses (9\,nm, 13\,nm, and 14\,nm), deposited on a 300\,nm thick thermal-silicon-oxide (SiO$_2$) on silicon substrate. The wire width $w$ was smaller than the Pearl length $\Lambda \approx 8.9$\,µm (estimated in Section 6 of the supplementary material for $d=14$\,nm) and, therefore, their vortex surface barriers followed the description of Nakagawa et al.\cite{nakagawa_vortex_2024} for narrow strips ($w\ll\Lambda$).
    Figure \ref{fig: 1}a shows the scanning electron micrograph of the device and an illustration of the operating principle.
    The triangular notch, larger than the London penetration depth $\lambda_\text{L}\approx250$\,nm and comparable to the wire width, lowers the vortex surface barrier on the left side due to current crowding \cite{hortensius_critical-current_2012}. At the critical current Lorentz forces push vortices from the notch towards the center of the wire, generating a hotspot (resistive state). When an out-of-plane magnetic field is applied, a screening current flows in opposite directions along the two edges of the wire (Meissner effect). For a positive bias current $I^+$ and field (directed into the plane), the screening current at the notch tip combines with the bias current, further suppressing the left-side barrier and reducing the critical current (the net Lorentz force increases). For a negative bias current $I^-$, the Meissner and bias currents oppose each other at the tip, increasing the barrier and raising the critical current. Inverting the field orientation inverts the direction of the circulating current, and thus reverses the surface barrier suppression, inducing the diode effect with opposite polarity. 
    Figure \ref{fig: 1}b shows the associated circuit symbols.
    The effect is explained in detail in Section 2 of the supplementary material, supported by time-dependent Ginzburg-Landau (TDGL) simulation results of the used geometry. 
    
    %diode geometry
    We optimized the diode by patterning 20 superconducting devices on a 14\,nm thick film with constant wire width (1\,µm) but varying constriction width and aperture angle of the triangular defect. The thickness was chosen to be comparable to typical values for superconducting nanowire electronics. 
    We observed the SDE in every device at 4.2\,K, with consistent diode polarity and a similar rectification efficiency $\eta$ (from 28\,\% to 37\,\%). The efficiencies for the 20 measured devices are reported in Section 3 of the supplementary material. The maximum value of $\eta$ was obtained using a notch with a $45^{\circ}$ angle, and a 400\,nm wide constriction (Figure \ref{fig: 1}a). 

    %bridge rectifier
    Bridge rectifiers are more power efficient than single-diode rectifiers in AC-to-DC current conversion because both halves of the input waveform are rectified (full-wave rectification).
    Therefore, such circuits are preferable for power distribution networks.    
    After rectification, the DC current is usually achieved by using a low-pass filter. 
    Figure \ref{fig: 1}c shows the circuit schematic of the superconducting bridge rectifier driven by an ideal current source and coupled to the low-pass filter. 
    Thanks to the duality between superconducting diodes and their conventional counterparts, the topology of the circuit is familiar. 
    The rectifier is composed of four vortex diodes ($D_1$, $D_2$, $D_3$, and $D_4$) connected in a superconducting loop.

    %full wave rectification
    We fabricated a bridge rectifier using the optimal diode geometry on the 13\,nm thick film.
    Firstly, we tested just the bridge circuit with a surface-mount resistor as a load, obtaining full-wave rectification of various input waveforms while using differential inputs and outputs to achieve isolation from the ground.
    We then added a smoothing capacitor to the output to demonstrate AC-to-DC conversion. 
    Figure \ref{fig: 1}d shows full-wave rectification of a 3\,MHz sinusoidal signal (without smoothing capacitor), for two values of the applied magnetic field. 
    The sign of the output current $i_\text{OUT}$ could be tuned by flipping the field, as all the diodes in the circuit had tunable polarity.

    \section{Rectification efficiency}\label{sec: char_diode}

\begin{figure*}[htbp!]
          
            \includegraphics[width=15.5cm]{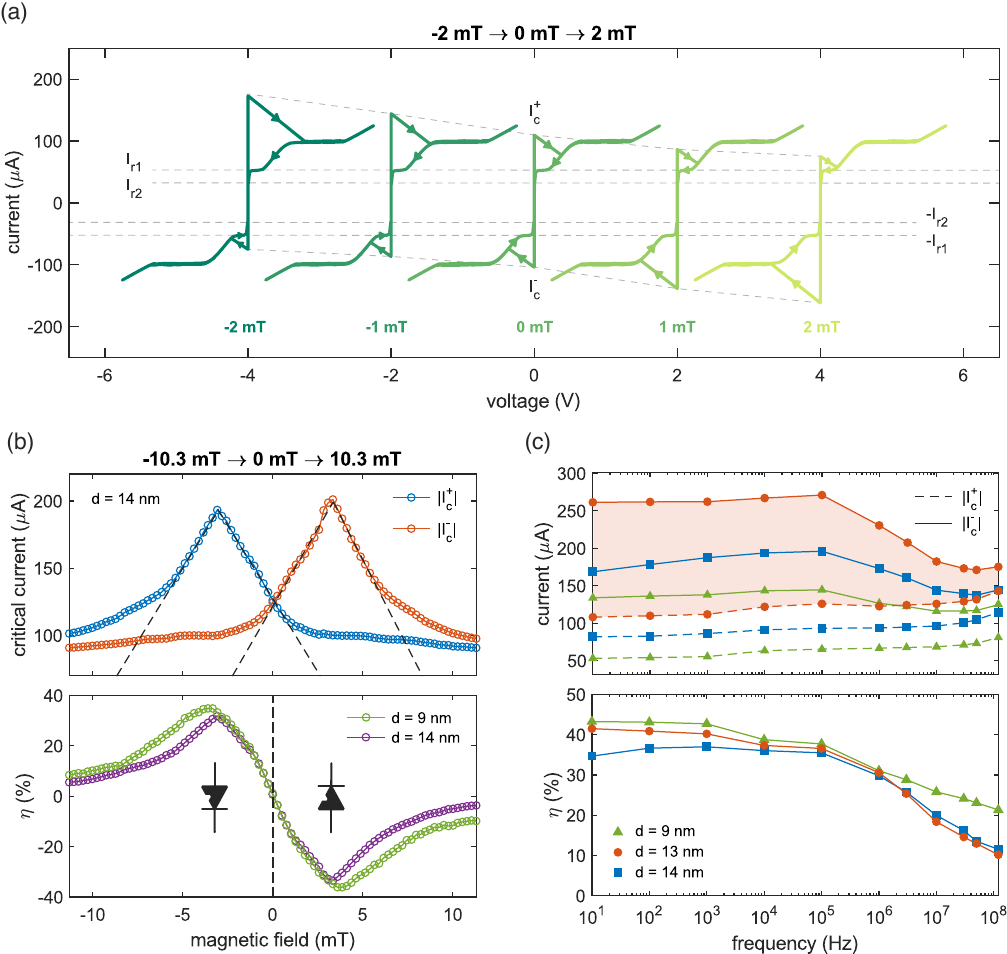}
            \caption{Characterization of superconducting diodes. (a) I-V curves of a superconducting diode, fabricated on a 14\,nm thick film, for different values of magnetic field. The field was swept from -2\,mT to 2\,mT. The curves are shifted along the x-axis for clarity. Positive and negative critical currents ($I^{+}_\text{c}$ and $I^{-}_\text{c}$) and two of the  retrapping currents ($I_{r1}$ and $I_{r2}$) are indicated by dashed lines. Arrows indicate the directions of current sweeps. (b) Upper panel: $|I^{+}_\text{c}|$ and $|I^{-}_\text{c}|$ as a function of the magnetic field (field swept from -10.3\,mT to 10.3\,mT). Dashed lines fit the linear regions of the plot. Lower panel: Rectification efficiency $\eta$ as a function of the magnetic field with a triangular wave as bias signal, for two different film thicknesses $d = 9, 14$\,nm and the same geometry. (c) Current margins to perform half-wave rectification (upper panel), and rectification efficiency $\eta$ (lower panel), for a sinusoidal bias signal at different frequencies. Devices with three different thicknesses $d = 9, 13, 14$\,nm were tested. For each $d$, the applied magnetic field was chosen to maximize $\eta$ at 10\,Hz. In the top panel, the data points on the dashed lines correspond to $|I^{+}_\text{c}|$ and data points on the full lines are associated with $|I^{-}_\text{c}|$. The colors and shapes of the data points of traces in both panels are associated with the legend. The highlighted red region is associated with $d=13$\,nm, thickness used for the bridge rectifier.}  \label{fig: 2} 
        \end{figure*}

        %IV curves:
        The rectification efficiency $\eta  = \frac{|I^{+}_\text{c}|-|I^{-}_\text{c}|}{|I^{+}_\text{c}|+|I^{-}_\text{c}|}$ is a critical figure of merit to be maximized for optimal half-wave and full-wave rectification. We measured the I-V characteristic of the diodes to extract the positive and negative critical currents ($I^{+}_\text{c}$ and $I^{-}_\text{c}$ respectively).
        Figure \ref{fig: 2}a shows the I-V curves of a diode patterned on the 14\,nm thick film, for different values of the field. With positive magnetic fields (entering the plane) the negative critical current $|I^{-}_\text{c}|$ increased, and the positive critical current $|I^{+}_\text{c}|$ decreased. Negative magnetic fields caused the opposite effect. 
        For three characteristic values of retrapping currents (or hotspot currents) \cite{berggren_superconducting_2018} stable hotspots could be sustained in sections of the wire having different widths. The largest value (large current plateau in Figure \ref{fig: 2}a) is associated with the micro-bridges that connected the device to the pads (3\,µm wide); $I_{r1}$ is for for the 1\,µm wide part of the diode; and the $I_{r2}$ is related to the hotspot in the 400\,nm wide constriction. Their values did not vary with the applied field: $I_{r1} = 52$\,µA, $I_{r2} = 28$\,µA for the 14\,nm thick film and $I_{r1} = 60$\,µA, $I_{r2} = 33$\,µA for the 13\,nm thick film, which was used for the bridge rectifier (a close-up view of the I-V curve is in Section 4 of the supplementary material).
    
        %Ic  and rectification factor
        We studied how the critical currents and rectification efficiency vary with the magnetic field by applying a 1\,kHz triangular wave to a device shunted with 1\,M$\Omega$.
        The upper panel of Figure \ref{fig: 2}b shows that $|I^{+}_\text{c}|$ and $|I^{-}_\text{c}|$ were equal at zero magnetic field. 
        $|I^{-}_\text{c}|$ linearly increased with an increasing positive field up to a peak value, while $|I^{+}_\text{c}|$ decreased. After the peak, $|I^{-}_\text{c}|$ linearly decreased because the surface barrier on the right edge became smaller than the barrier on the left.  
        At about $\pm 5.5$\,mT, the relation ceased to be linear and both negative and positive currents approached similar low values because bulk vortex pinning effects dominated \cite{suri_non-reciprocity_2022}. 
        The peak of $|I^{-}_\text{c}|$ was higher than the peak of $|I^{+}_\text{c}|$ (2\,\% of the peak current), as observed in similar works \cite{suri_non-reciprocity_2022, hou_ubiquitous_2024}. A possible explanation is that the defect on the right edge, which influenced the current density on that side, varied depending on the current's direction. We observed a 2.5\,\% asymmetry for the 9\,nm thick film (shown in Section 5 of the supplementary material). 

        From the slope $s=22.5$\,µA/mT ($s=8$\,µA/mT for the 9\,nm film) of the $|I^{-}_\text{c}(B)|$ curve between 0\,mT and 3.3\,mT, we could extract the London penetration depth \cite{kuit_vortex_2008, plourde_influence_2001, vodolazov_superconducting_2005}:  $\lambda_\text{L} \approx 1.22\sqrt{dw^2/2\mu_0s}\approx 250$\,nm and Pearl length $\Lambda = 2\lambda^2_\text{L}/d\approx 8.9$\,µm. $d$ is the thickness, $w$ is the constriction width, and 1.22 is a correction factor we estimated with TDGL simulations to account for the asymmetric current distribution in the wire (see Section 6 in the supplementary material). 
        The values of $\lambda_\text{L}$ are consistent with those reported in literature \cite{medeiros2022investigation, luo_niobium_2023}.
        From the intercepts of the $|I_\text{c}(B)|$ linear fits with the x-axis between the two current peaks, we found the reduced critical field for vortex entry defined by the notch: $b_\text{s}\approx5.6$\,mT for $d=14$\,nm. This value is comparable to our theoretical estimation \cite{buzdin_electromagnetic_1998,maksimova_mixed_1998}  $b_\text{s} \approx (2\xi/\Lambda)^{3/7} \phi_0/2\pi\xi w\approx5.7$\,mT, where the coherence length $\xi = 10.8$\,nm was calculated from Ginzburg-Landau theory. With $\xi = 5$\,nm, a value closer to experimental results for NbN in literature \cite{luo_niobium_2023}, $b_\text{s}$ is 8.9\,mT. 
        More details on these estimations are in Sections 1 and 7 of the supplementary material, where we also discuss the critical field $B_\text{s}$ defined by the right edge and the ratio $b_\text{s}/B_\text{s}$ for different $d$.

        The lower panel of Figure \ref{fig: 2}b shows the rectification efficiency $\eta$ as a function of the magnetic field for two different film thicknesses (14\,nm and 9\,nm). For the 14\,nm thick device the maximum rectification efficiency of 35\,\% was reached at 3.3\,mT. The 9\,nm thick device had a maximum $\eta$ of 36\,\% at 4\,mT.

        %rectification factor as a function of frequency
        Demonstrating signal rectification at frequencies in the MHz range would enable radio-frequency signals to be processed and converted into DC levels with lower circuit footprints, and would thus be enabling for the future development of this technology. Therefore, we studied how the rectification efficiency varies as a function of the frequency of an applied sinusoidal signal. 
        Figure \ref{fig: 2}c shows the values of $|I^{-}_\text{c}|$, $|I^{+}_\text{c}|$, and $\eta$ at frequencies between 10\,Hz and 120\,MHz for three different film thicknesses (with a 50\,$\Omega$ load).
        For all the thicknesses, $\eta$ started substantially decreasing at frequencies higher than 100\,kHz (a similar cutoff has been observed by Chahid et al. \cite{chahid_high-frequency_2022}). 
        Below 1\,MHz, $\eta$ increased with thinner films, as predicted by Vodolazov et al. \cite{vodolazov_superconducting_2005}.
        Above 1\,MHz, the 13\,nm and 14\,nm  thick films started decreasing with a faster slope than the 9\,nm thick film: by $\sim 20\,\%$ instead of $\sim 10\,\%$ between 1\,MHz and 120\,MHz, and $\eta(d=13\,\text{nm})$ dropped below $\eta(d=14\,\text{nm})$ around 3\,MHz. A possible cause of this behavior is explained in Section 11 of the supplementary material.
        The highest rectification efficiency of 43\,\% was achieved at 10\,Hz, for the 9\,nm thick film.
        At the same frequency (1\,kHz) and same thickness (9\,nm or 14\,nm), we observed different maxima of $\eta$ for sinusoidal signals with a 50\,$\Omega$ shunt (Figure \ref{fig: 2}c) and triangular waves with a 1\,M$\Omega$ shunt (Figure \ref{fig: 2}b). This result shows that shape of the signal and shunt impedance influence the efficiency.

        %half-wave rectification
        We observed half-wave rectification with a single diode when the applied current amplitude was between $|I^{+}_\text{c}|$ and $|I^{-}_\text{c}|$.
        The 13\,nm thick film had the largest critical currents and margins (region highlighted in red). With this film, we rectified sinusoidal signals at 50\,MHz and 120\,MHz with a ratio between output and input amplitudes of 0.63 and 0.54 respectively (see Section 9 in the supplementary material). 

    \newpage
       
    \section{Margins for full-wave rectification}\label{sec: char_bridge}

\begin{figure*}[htbp!]
          
            \includegraphics[width=15.9cm]{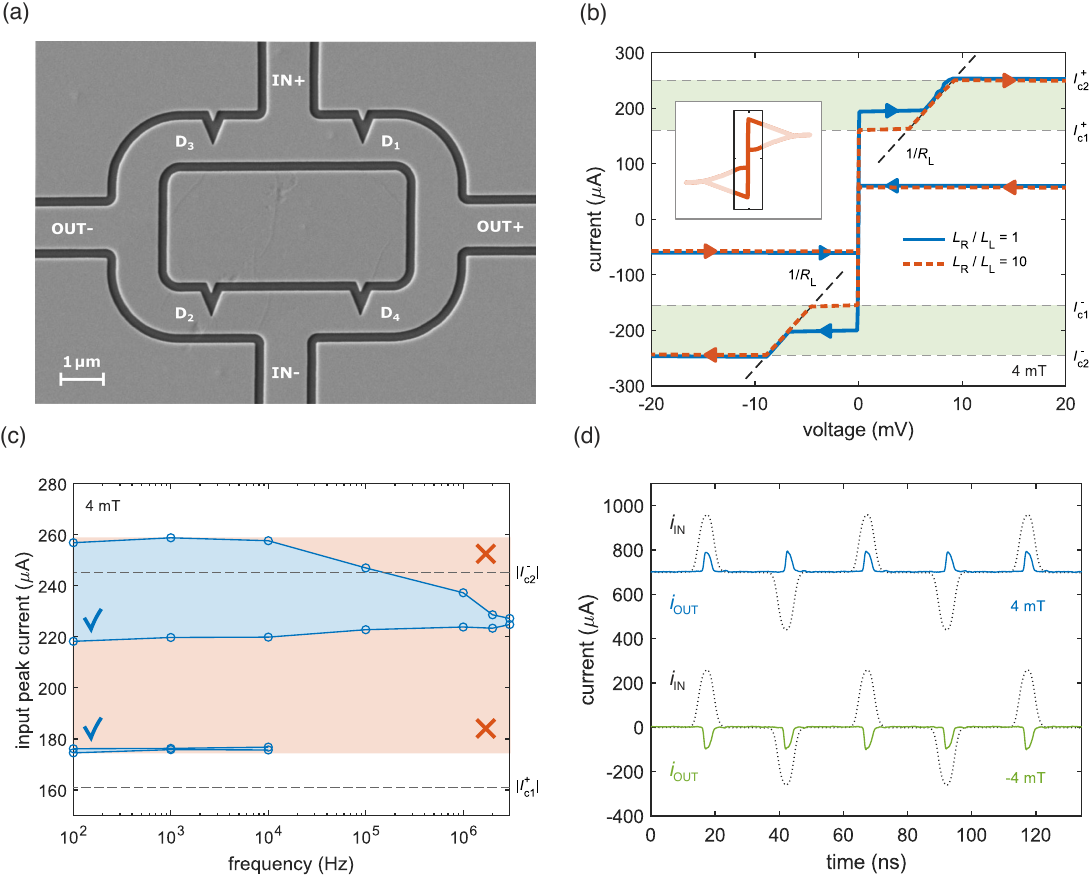}
            \caption{Characterization of superconducting bridge rectifiers. (a) Scanning electron micrograph of a superconducting bridge rectifier, fabricated on a 13\,nm thick film of niobium nitride ($L_\text{R}/L_\text{L}=1$). The gray regions are niobium nitride, and the dark borders of the traces show the underlying SiO$_2$ substrate. The ground plane surrounds all the traces. The current source connects the IN- port to the IN+ port. The load resistor $R_\text{L}$ is connected between OUT- and OUT+ (see Figure \ref{fig: 1}c). The geometry of the rectifier with $L_\text{L} / L_\text{R} = 10$ is shown in the supplementary material. (b) Close-up view of I-V curves for bridge rectifiers with two different values of $L_\text{R}/L_\text{L}$, with a 4\,mT field ($R_\text{L} = 50$\,$\Omega$). The full curve is shown in the top left inset. Green regions indicate ideal current margins for the correct operation of the rectifier with $L_\text{R}/L_\text{L}=10$. (c) Margins to obtain correct full-wave rectification, for the input current amplitude of a periodic sinusoidal signal $i_\text{IN}$, as a function of frequency (field: 4\,mT). The two dashed lines indicate the ideal minimum and maximum currents to operate the rectifier according to figure b ($L_\text{R}/L_\text{L}=10$). In blue regions, the device correctly rectifies. In red and white regions, the device does not rectify. (d) Full-wave rectification of periodic current pulses at 20\,MHz, for positive and negative magnetic fields. The measured input signals $i_\text{IN}$ are shown with dashed lines. Traces are vertically shifted for clarity. Each trace is the average of 100 different traces acquired in sequence, to increase the signal-to-noise-ratio. An output pulse was observed for every semi-period.}\label{fig: 3}
        \end{figure*}

        Figure \ref{fig: 3}a shows a superconducting bridge rectifier patterned on the 13\,nm thick film. The left branch of the loop is composed of $D_2$ and $D_3$, the right branch is composed of $D_1$ and $D_4$. Both branches had a characteristic total kinetic inductance ($L_\text{L}$ and $L_\text{R}$ respectively), and the ratio $ L_\text{R} / L_\text{L}$ influenced the operations and margins of the device. We fabricated and characterized bridge rectifiers with $L_\text{R} / L_\text{L}=1$ and $L_\text{R} / L_\text{L}=10$. The device with $L_\text{R} / L_\text{L}=10$ is shown in Section 12 of the supplementary material.

        %I-V curve
        Figure \ref{fig: 3}b shows the I-V curves of two devices with different $L_\text{R} / L_\text{L}$, while the applied positive field (4\,mT) was set to maximize $\eta$. For both devices, we observed two switching events at positive and negative currents: $I^{+}_\text{c1}$, $I^{+}_\text{c2}$, $I^{-}_\text{c1}$, and $I^{-}_\text{c2}$.
        For the positive quadrant of the plot, when the current through $D_3$ and $D_4$ was lower than their $I^{+}_\text{c}$ ($i_\text{IN}<I^{+}_\text{c1}$), the input current $i_\text{IN}$ was split into the two branches. Thus, no hotspots were generated and no current was provided to the load resistor (zero voltage in the I-V curve). 
        In this region, for $L_\text{R} / L_\text{L}=1$, the current was equally split. For $L_\text{R} / L_\text{L}=10$, most of the current flowed into the left branch ($i_\text{IN}/(1+L_\text{L} / L_\text{R})$).
        When the current through both $D_3$ and $D_4$ exceeded their $I^{+}_\text{c}$, $D_3$ and $D_4$ switched to the resistive state, and most of the input current was diverted to $D_1$ and $D_2$, which remained in the superconducting state. Therefore, most of $i_\text{IN}$ passed through the load resistor, and the slope in the I-V curve was $1/R_\text{L}$ ($I^{+}_\text{c1}<i_\text{IN}<I^{+}_\text{c2}$). The current through the load was $i_\text{OUT} \approx  i_\text{IN}-2I_\text{r}$, where $I_\text{r}$ was a function of load resistance and input current according to the I-V curve in Figure S4 of the supplementary material ($I_\text{r}\ge I_\text{r2}$). When the currents through $D_1$ and $D_2$ exceeded $I^{-}_\text{c}$ ($i_\text{IN}>I^{+}_\text{c2}$), all the diodes switched, resulting in a high resistance state in the I-V curve. Then, all the hotspots retrapped during the downward sweep when the current was approximately $2I_\text{r2}$. 
        When the input current was negative, $D_3$ and $D_4$ were superconducting and $D_1$ and $D_2$ were resistive so that the output current could flow in the same direction as with positive $i_\text{IN}$.

        % The current through the load was $i_\text{OUT} \approx  i_\text{IN}-2I_\text{r}$, where $I_\text{r}$ is a function of load resistance and input current according to the I-V curve of Fig. S4 in the supplementary material ($I_\text{r}\ge I_\text{r2}$)
        
        %ideal operation range 
        For ideal full-wave signal rectification, the device should operate with a positive peak input current between $I^\text{+}_\text{c1}$ and $I^\text{+}_\text{c2}$, and a negative peak current between $I^\text{-}_\text{c2}$ and $I^\text{-}_\text{c1}$ in Figure \ref{fig: 3}b. 
        The switching currents, and consequently the margin of the device, are influenced by the inductance ratio $r=L_\text{L} / L_\text{R}$, which affects the diode switching dynamics and the generation of a parasitic circulating current $I_\text{circ}(r)$ (positive if counter-clockwise) in the superconducting loop. This circulating current can arise when the two branches retrap either simultaneously or at different times in the preceding measurement (or period), depending on $r$ and small variations in the retrapping currents $I_\text{r2}$ within the circuit.
        If $r>[I_\text{r2}+I_\text{circ}(r)]/[|I^{+}_\text{c}|- I_\text{circ}(r)]$, a condition that holds for $r=1$ ($I_\text{circ}(0)\approx0$\,µA), after $D_3$ switches in the upward sweep, sufficient current is diverted to the right branch, causing $D_4$ to rapidly switch as well. Under this condition, the switching currents are $|I^\text{+}_\text{c1}| = |I^\text{-}_\text{c1}| \approx (1+r)|I^{+}_\text{c}|$ and $|I^\text{+}_\text{c2}| = |I^\text{-}_\text{c2}| \approx |I^{-}_\text{c}|+I_\text{r}$ with $I_r\ge I_\text{r2}$.
        Conversely, if the condition on $r$ is not satisfied, the current diverted to the right branch is insufficient to switch $D_4$, causing $D_3$ to reset with approximately $I_{r2}$ flowing through it. The current then continues increasing until $D_3$ switches again, and the diverted current switches $D_4$ as well (double-switch behavior). 
        In this scenario, valid for $r=0.1$, the lower switching currents become 
        $|I^\text{+}_\text{c1}| = |I^\text{-}_\text{c1}| \approx (1+r)[2|I^{+}_\text{c}|- I_\text{r2}-I_\text{circ}(r)]$.
        Both the double-switch behavior and the circulating current generation are explained in detail in Sections 14 and 15 of the supplementary material.
        
        For both circuits, the observed retrapping current $I_\text{r2}$ was approximately 30\,µA and the extracted $I^-_\text{c}$ was 220\,µA, which is lower than the value in Figure \ref{fig: 2}c. This difference might be attributed to smaller constrictions or variations in the optimal field values for the diodes within the loop. 
        For $r=1$, $I^+_\text{c}=|I^\text{+}_\text{c1}|/2 = 100$\,µA was comparable to the value in Figure \ref{fig: 2}c (108\,µA). 
        For $r=0.1$, we estimated $I_\text{circ}(r)\approx I_\text{r2}(1-r)/(1+r)\approx25$\,µA by assuming the two branches retrapped simultaneously (more details in Section 15 of the supplementary material) and thus $|I_\text{c1}|\approx 168$\,µA with $I^+_\text{c}\approx104$\,µA, average between 100\,µA and 108\,µA. The value of $|I_\text{c1}|$ is in agreement with the experiment ($|I^{+}_\text{c1}|=161$\,µA, $|I^{-}_\text{c1}|=155$\,µA). 
    
        %operation range as a function of frequency
        We further characterized the rectifier with the largest margins ($L_\text{R}/L_\text{L}=10$) to study how they varied as a function of the frequency of a sinusoidal signal. 
        Figure \ref{fig: 3}c shows the range of $i_\text{IN}$ amplitudes (the region in blue) that allowed full-wave rectification, for different frequencies. 
        Outside of these margins, either only half-wave rectification was observed, or the output current was zero. 
        The red region is bounded by the minimum and maximum $i_\text{IN}$ values where correct operation was observed in the measured spectrum. 
        Theoretically, these should be $|I^{+}_\text{c1}|=161$\,µA and $|I^{-}_\text{c2}|=245$\,µA from Figure \ref{fig: 3}b, but both showed a positive shift of about 15\,µA, likely due to differences in the measurement setups or hotspot retrapping dynamics.   
        Additionally, the margins were narrower than expected from the I-V curve
        In particular, they were discontinuous between 100\,Hz and 10\,kHz: for example, at 100\,Hz, the device worked properly between 175\,µA and 176\,µA; it did not operate correctly from  176\,µA to 218\,µA; and it rectified again between 218\,µA and 257\,µA.
        Above 10\,kHz the margins were continuous but gradually reduced from 38\,µA at 10\,kHz to 3\,µA at 3\,MHz. 
        The 3\,MHz frequency limit was caused by the decrease of $\eta$ and $|I^{-}_\text{c}|$ in the diode, observed in Figure \ref{fig: 2}c ($\eta\approx26$\,\% and $|I^{-}_\text{c}|\approx208$\,µA at 3\,MHz). In particular, the top margin, approximately $I^{-}_\text{c}+I_\text{r}$, decreased with $I^{-}_\text{c}$.
        
        Based on the calculations in Section 15 of the supplementary material, rectification above 218\,µA could be associated with the most likely scenario of the left branch completely retrapping before the right one (due to the left lower impedance), for which we estimated $I_\text{circ}(r)\approx-3rI_\text{r}/(1+r)\approx-8$\,µA, and thus correct rectification only above $|I_\text{c1}|\approx205$\,µA at 100\,Hz and above $|I_\text{c1}|\approx230$\,µA at 3\,MHz ($I^+_\text{c}(3\,\text{MHz})=115$\,µA).  
        The unstable behavior observed around 175\,µA may correspond to a scenario where the two loop branches retrap at two close times. Indeed, for diodes resetting simultaneously in this measurement, we estimated $|I_\text{c1}|\approx155$\,µA.
        These estimations are comparable to the experimental margins.

        %%%%%%%%%%%%%%%%%%%
            
        % amplitude ratio
        Maximizing the ratio between the peak output current and input current amplitude $I_\text{OUT}/I_\text{IN}$ is fundamental for optimizing the power conversion efficiency of the rectifier. With a 225\,µA input amplitude at 3\,MHz, its value was 0.76, approximately equal to $1-2I_\text{r}/I_\text{IN}$, with $I_\text{r}\approx I_\text{r2}$. In general, $I_\text{r}$ and the ratio can change with input amplitude and load impedance.
        
        % pulse shape
        Figure \ref{fig: 3}d shows that the circuit could also correctly rectify 10\,ns wide pulses at a maximum frequency of 20\,MHz, but with a lower efficiency than for a sinusoidal signal at 3\,MHz: the peak output current was about 3 times smaller than the input peak current.
        The higher operating speed with pulses is consistent with the idea that the duty cycle of the waveform, and thus the heating of the substrate, is critical to correct operation. 
        
          %model        
        We created a simple LTspice model of the vortex diode, based on the electrothermal model of NbN nanowires \cite{berggren_superconducting_2018}, to better describe the experimental behaviors of the bridge rectifier.
        In simulations, we mimicked full-wave rectification at 3\,MHz and observed the effects of the inductance ratio to the margins: generation of circulating current and double-switching behavior (see Section 14 in the supplementary material).

\section{AC-to-DC conversion}\label{sec: ACDC}

\begin{figure*}[htbp!]
          
            \includegraphics[width=16cm]{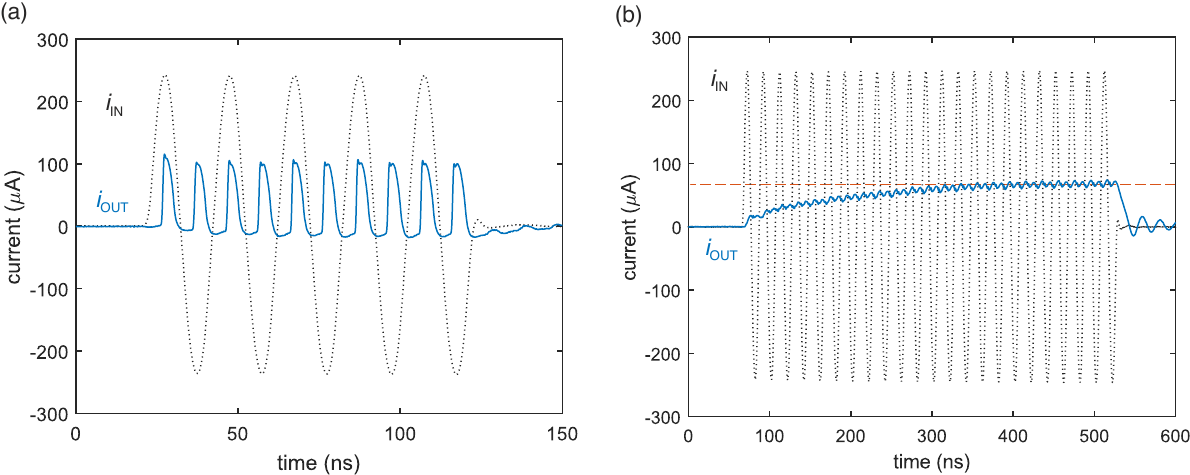}
            \caption{Operating the bridge rectifier and AC-to-DC converter at 50\,MHz. (a) Full-wave rectification of a sinusoidal signal for 100\,ns. The input signal $i_\text{IN}$ is set to burst mode with a 1\,µs burst period. The load $R_\text{L}$ is 50\,$\Omega$ (100\,$\Omega$ surface-mount resistor in parallel with the 100\,$\Omega$ differential mode impedance from the amplifiers). There is a transient for the lower level of the output signal $i_\text{OUT}$ because the input capacitance of the used amplifiers need to be charged. (b) AC-to-DC conversion with a 500\,pF smoothing capacitor (see Figure \ref{fig: 1}c), for 440\,ns in burst mode (the burst period is 100\,µs). The red dashed line indicates the value of the output DC current (69\,µA) between 340\,ns and 520\,ns. Each waveform is the average of 100 different traces acquired in sequence, to increase the signal-to-noise-ratio.} \label{fig: 4}
        \end{figure*}

        %rectification at higher frequencies
        The device could rectify continuous sinusoidal signals up to 3\,MHz.
        However, we observed rectification at higher frequencies when the signal was applied in short periodic bursts (burst mode).
        Figure \ref{fig: 4}a shows a fully rectified 50\,MHz sinusoidal signal applied for 100\,ns with a 1\,µs burst period.
        At this frequency, a longer burst duration caused instability in the system (only a portion of semi-periods were rectified after 100\,µs). 
        We believe the observed difference in maximum operating frequency between the continuous and burst modes was due to the system heating up significantly during the continuous operation while having time to cool off between bursts in burst mode. We investigated the over-heating effect by studying how the temperature affects the critical currents of a single diode, observing a strong dependence of $|I^-_{c}|$ ($\Delta I \approx- (50\,\text{µA}/\text{K}) \Delta T$). Afterward, we measured how $|I^-_{c}|$ and $|I^+_{c}|$ of the diode evolve in time for a burst signal at 3\,MHz, 10\,MHz, and 50\,MHz, finding fast and large critical-current decays in time. We found a correlation between the current values of the two measurements, suggesting that the local temperature increased during the burst and Joule heating was the dominant effect limiting $\eta$. The $\eta$ decrease over time was larger for higher frequencies, in agreement with the frequency dependence of $\eta$ in Figure \ref{fig: 2}c.  These results, shown in Section 11 of the supplementary material, suggest that the rectifier stopped operating after 100\,ns because the margins became smaller than the input current.       
        
        %rectification with capacitor 
        Considering that the circuit could operate up to 50\,MHz in burst mode, we demonstrated AC-to-DC conversion by shunting the load $R_\text{L}$ with a 500\,pF surface-mount smoothing capacitor (dashed line in Figure \ref{fig: 1}c) as a proof of principle. 
        Figure \ref{fig: 4}b shows the output current of the converter through the load resistor.
        The device could operate for a 460\,ns long burst (burst period: 100\,µs), with an 8\% ripple whose amplitude was set by the capacitance $C$ (A larger $C$ can be used to decrease the ripple amplitude or operate a lower frequency). The DC value of the current saturated to 69\,µA. The waveform in the figure was averaged over 100 signals to increase the SNR and thus extract the ripple amplitude. After a start-up time of about 300\,ns, the capacitor was correctly charged at each semi-period, for all the acquired signals.
        
        It is worth mentioning that including the smoothing capacitor enabled correct rectification for a longer burst duration compared to the configuration with a resistive load. 
        This difference likely comes from the intrinsic dependence on load impedance for the hotspot growth and, consequently, for thermal dissipation.

    %short intro on LTspice modeling and power efficiency
    We could not experimentally extract the power efficiency of the circuit because sensing the input voltage would have altered the operation of the device. Therefore, we estimated the efficiency by mimicking the AC-to-DC conversion with our LTspice model of the diode end extracting the value from the simulation.
    The estimated peak power efficiency, defined as the ratio between power dissipated in $R_\text{L}$ and total power consumption (450\,nW without considering bias resistors) after the start-up time, was about 50\,\% with a 50\,$\Omega$ load and a 500\,pF smoothing capacitor. 
    Figure S15 in the supplementary material shows the comparison between the simulation result and a single experimental trace without averaging.

\section{Conclusions}\label{sec: dis}

    We have successfully demonstrated a superconducting circuit based on vortex diodes, showing correct full-wave rectification and characterizing the device margins. 
    As a proof-of-concept, we achieved efficient AC-to-DC conversion by shunting the resistive load with a surface-mount smoothing capacitor to obtain a low-pass filter. 
    However, integrated capacitors would increase the footprint of the device which is undesirable.
    Instead, in future work the filter could be designed using inductors rather than capacitors, exploiting the high kinetic inductance of superconducting nanowires.

    In Section 17 of the supplementary material, we explored using an AC-to-DC converter with an inductor-based filter to create a bias distribution network for SNSPDs.
    In our analysis, we had to assume that the diodes were optimized to operate at 50\,MHz and higher frequencies in continuous mode with ideal margins (and thus this design is conditional on significant future advancement of the work demonstrated experimentally so far).
    Nonetheless, the design demonstrates the relevance and future potential of superconducting diodes in any cryogenic system that requires dynamically tunable DC bias currents. 
 
    The operational speed of our devices was mainly limited by Joule heating.
    Considering that relaxation oscillations in NbN superconducting nanowires have been observed up to 1.25\,GHz \cite{castellani_design_2020}, it may be possible to optimize geometry, material properties of the diode, and the load on the device to minimize the hotspot growth and heat dissipation and reach higher frequencies (e.g. by changing substrate to improve thermal transfer).
    Moreover, new circuit topologies that limit the generation of circulating currents might be introduced to enlarge the operating margins, and thus the tunability range for the output current.
    In Section 11 of the supplementary material, we estimate the optimal inductance ratio that maximizes the margin according to our experimental results ($r\approx0.15$).

    Improving the rectification efficiency will benefit margins, speed, and device footprint. 
    Following our results and the literature \cite{vodolazov_superconducting_2005}, it might be possible to further improve $\eta$ by decreasing the thickness to 5\,nm or less (typical for SNSPDs).  
    In particular, bridge rectifiers on a 9\,nm thick film should be tested in the future, considering that we observed a higher $\eta$ for such a thickness (larger wires should be used to keep the same signal-to-noise ratio).
    Furthermore, device efficiency could be enhanced by operating at lower temperatures, as our measurements revealed a significant temperature dependence of $\eta$ ($\Delta\eta\approx-0.089\,\text{K}^{-1}\Delta T$ in Section 11 of the supplementary material).
    We fabricated only wires with a 1\,µm width (smaller than the Pearl length), but using wider wires might also increase $\eta$.
    In a wider micro-bridge under the same applied magnetic field, higher Meissner currents are generated and thus the asymmetric effect might increase.  
    Moreover, if the width exceeds the Pearl length of the film, the current flows along the edges, and thus closer to the defect which might further improve the efficiency of the diode.

    %External B-field
    For future designs of large-scale diode circuits, the need for an external magnetic field uniformly applied across the chip will limit the integration with field-sensitive superconducting devices. 
    Therefore, vortex diodes based on NbN could be integrated with ferromagnetic thin films to localize the field and enhance the rectification efficiency, similar to the approach by Hou et al. \cite{hou_ubiquitous_2024}.
    
    The results of this work can be seen as a fundamental starting point for future development in the field of integrated power electronics based on superconducting diodes in thin film platforms.

    \textit{Note added}: during the submission process of this work we became aware of a related work
on superconducting rectifiers \cite{inglaaynés2024highlyefficientsuperconductingdiodes}.

\section{Methods}\label{sec: met}

    \subsection{Fabrication process}\label{sec: fab}

        All the tested devices were fabricated with the same process: we deposited the NbN films on 300\,nm thick silicon oxide on silicon substrates in an AJA sputtering system; then, we patterned the devices with electron-beam lithography, using a positive-tone resist (ZEP530A) with cold development in o-Xylene, and CF$_4$ reactive ion etching (RIE) \cite{castellani_nanocryotron_2024}.

    \subsection{Measurement setup for I-V curves}\label{sec: meas_IV}

        %measurement with dewar
        We performed the measurements in a liquid helium dewar at 4.2\,K. The chip was glued to a PCB and the pads of the devices were connected to the pins through aluminum wire bonds. The PCB was then placed in a custom cryogenic probe with 28 spring-loaded RF sub-miniature push-on connectors \cite{butters_digital_2022}. 
        The out-of-plane magnetic field was applied with a superconducting solenoid mounted on the cryogenic probe in close proximity to the chip. The solenoid was controlled by a DC current source.

        %IV curves

        %critical current vs field and frequency and half-wave rectification
        In Figure \ref{fig: 2}b, the critical currents for each data point were found by averaging over 500 I-V curves, which were obtained by applying a 1\,kHz triangular waveform with an arbitrary waveform generator (AWG, Agilent AWG33622A). A 10\,k$\Omega$ bias resistor was used and the voltage across the diode was measured with a 2\,GHz real-time oscilloscope (LeCroy 620Zi) with 1\,M$\Omega$ input impedance.
        For frequencies between 10\,Hz and 100\,kHz in Figure \ref{fig: 2}c, we used the same configuration but the impedance of the oscilloscope was set to 50\,$\Omega$, and the biasing signal was a sinusoidal wave. Each point of the curve is the average of 1000 values.
        For both Figure \ref{fig: 2}b and Figure \ref{fig: 2}c (up to 100\,kHz), we calculated the average critical current $|I_\text{c}|$ and relative standard deviation (RSD) at every point, by fitting the data with Gaussian distributions. In Figure \ref{fig: 2}b, the average value of RSD is 2.7\,\% for $d = 9$\,nm, and 1.2\,\% for $t=14$\,nm.
        In Figure \ref{fig: 2}c, the average RSD is 1.9\,\% for $d = 9$\,nm, 1.4\,\% for $d = 13$\,nm, and 2.8\,\% for $d = 14$\,nm.

    \subsection{Measurement setup for half-wave and full-wave rectification}\label{sec: meas_bridge}

        We performed the measurements in a liquid helium dewar at 4.2\,K, as described in the previous Section.
        We used the same experimental setup to obtain the data between 1\,MHz and 120\,MHz in Figure \ref{fig: 2}c, the half-wave rectification of Figure S8 in the supplementary material, and all the full-wave rectification results. The input current $i_\text{IN}$ was provided with a differential method: the signal was sent to the IN+ terminal by the AWG in series with a 5\,k$\Omega$ surface-mount resistor placed on the PCB at 4.2\,K; simultaneously, the inverted signal was sent to IN- using the same configuration. The two input lines were 50\,$\Omega$ matched thanks to a surface-mount 50\,$\Omega$ resistor to ground in front of the bias resistors. A 100\,$\Omega$ load resistor was placed on the PCB at $4.2$\,K, and wire bonded to the OUT- and OUT+ terminals of the rectifier. For Figure \ref{fig: 2}c and S8, the OUT+ and IN+ ports correspond to one terminal of the diode, while the OUT- and IN- ports correspond to the opposite terminal (). 
        The output current $i_{\text{OUT}}$ was calculated by measuring the voltage across the load resistor $V($OUT+$)-V($OUT-$)$ with a differential readout: for Figure \ref{fig: 1}d, \ref{fig: 2}c \ref{fig: 3}c, \ref{fig: 3}d, \ref{fig: 4}b, and S8,  both the nodes were connected to the 50\,$\Omega$ input ports of the scope; for Figure \ref{fig: 4}a, both output voltages were filtered by a bias tee, attenuated by 10\,dB, and amplified by two low-noise amplifiers in series (RF Bay LNA-2500, bandwidth: 10 kHz to 2500 MHz, gain: 25 dB, and RF Bay LNA-2000, bandwidth: 10 kHz to 2000 MHz, gain: 26 dB). 
        In both configurations, the total load resistance $R_\text{L}$ was calculated considering the 100\,$\Omega$ resistor in parallel with the 100\,$\Omega$ differential resistance of the scope (or amplifiers).
        More detailed circuit schematics of the differential measurement setups are shown in Section 10 and 13 of the supplementary material. 
        For all the figures, the shown waveforms were obtained by averaging multiple traces.
        
        %margins measurement
        For Figure \ref{fig: 2}c and \ref{fig: 3}c, at each tested frequency, we swept the input current amplitude (from 100\,µA to 300\,µA in Figure \ref{fig: 3}c) with an increment of 0.2\,µA and observed if correct rectification was achieved. 
        To extract the critical currents in Figure \ref{fig: 3}c, we found the current amplitude for which an output voltage pulse was observed on average in half of the 100 sampled periods of the signal.  
        Data points in Figure \ref{fig: 3}c are associated with the current amplitude for which there was a transition between completely correct and incorrect operation.

    \subsection{LTspice modeling}\label{sec: LTspice}

        We created the LTspice model of the diode by slightly modifying the existing hotspot-growth model of a superconducting nanowire \cite{berggren_superconducting_2018}.
        We kept the same material parameters, modified the thickness and width, and introduced a critical current that depends on the current flow direction. 
        We did not model the hotspot growth in the real geometry of the device. Instead, we considered the diode as a 7.5-squares long wire with a constant width equal to the constriction width (400\,nm), and we varied the Steaky parameter to obtain a similar hotspot current (or retrapping current) as in the experiments. 
        For simulating the bridge rectifier, we extracted the number of squares for each branch from the device geometry (15 for the left branch and 150 for the right branch) to estimate $L_\text{R}$ and $L_\text{L}$. The sheet inductance was set to 30\,pH/$\square$ (see Section 12 in the supplementary material for more details). A lumped inductors was added in series with each diode to reach the values of $L_\text{R}$ and $L_\text{L}$.

        To model the diode limitations in frequency and margins as the experiment, the values of critical currents were set to the ones observed in Figure \ref{fig: 3}c, according to the frequency of the input signal. Therefore, the model did not recreate the dynamic evolution of $\eta$ due to substrate temperature changes caused by self-heating effects, shown in Figure S10 of the supplementary material. 
        We successfully modeled the time-domain behavior of the bridge rectifier shown in Figure \ref{fig: 1}d by setting the high critical current and $\eta$ to the corresponding values obtained in Figure \ref{fig: 2}c for the 13\,nm thick film at 3\,MHz (200\,µA and 26\,\% respectively). The currents were scaled by a factor of 104/108 to consider the results from the I-V curve of the rectifier. The retrapping current was set to 25\,µA to match the experimental output amplitude. 
        The comparison between the model and experimental data is shown in Section 14 of the supplementary material. 
        
        Considering that the power dissipation mainly depends on the modeled electrothermal behavior of the wires, we could estimate the power efficiency of the AC-to-DC converter by setting the rectification factor to 23\,\%  and $I^-_\text{c}$ to 240\,µA, to match the experimental output waveform. In a more accurate model that considers Joule heating and additional frequency dependencies of $\eta$, these values would change with time.
        In the supplementary material, we show the comparison between experimental and simulated output current at 50\,MHz in burst mode.
        The peak power efficiency was calculated after the start-up time when the output current saturated.

\section{Data availability}

The data that support the findings of this study are available within the
paper and the supplementary material. Additional data are available from the corresponding authors upon reasonable request.

\bibliography{Diode_paper}

\section{Acknowledgments}

This material is based upon work supported by the U.S. Department of Energy, Office of Science, Office of Basic Energy Sciences, under Award Number DE-AC02-07CH11359. 
O.M. acknowledges support from the NDSEG Fellowship program. 
A.B. acknowledges support from Politecnico di Torino.
R.A.F. acknowledges support from the DOE under the National Laboratory LAB 21-2491 Microelectronics grant.
M.Colangelo acknowledges support from MIT Claude E. Shannon award. 
The authors would like to thank Prof. Christoph Strunk and Prof. Philip Moll for advising in the interpretation of the results.

\section{Author contributions}

M.Castellani conceived the idea, fabricated the devices, supervised the experiment, analysed the data, and performed the simulations.
A.B. and O.M contributed to the design of the diode geometry.
A.B. O.M and R.A.F assisted in the measurement.
All the authors contributed to the discussions and production of the manuscript.
  
\nolinenumbers

\end{document}

\supplementarysection

\section{Theoretical estimation of film parameters}
   
Coherence length, London penetration depth, and Pearl length are fundamental parameters for modeling device physics.
We estimated the Ginzburg-Landau coherence length using $\xi = \sqrt{\hbar/\rho_\text{N} N_\text{F} e^2 \Delta_0 }$ and the London penetration length with $\lambda_\text{L} = \sqrt{\hbar\rho_\text{N}/ \pi \mu_0 \Delta_0 }$, where $\rho_\text{N} = R_{\square}d$ and $2\Delta_0 = 4.05k_\text{B}T_c$ (Bardeen–Cooper–Schrieffer energy gap).
$N_\text{F} \approx 10^{28}\,m^{-3}eV^{-1}$ is the density of states in NbN at the Fermi level \cite{suri_non-reciprocity_2022}.
For the 14\,nm thick film ($T_c = 8.6$\,K, $R_{\square}=170\,\Omega/\square$) we obtained $\xi \approx 10.8$\,nm and $\lambda_\text{L} \approx 514$\,nm, resulting in a Pearl length $\Lambda = 2\lambda^2_\text{L}/d$ of 38\,µm.
For the 9\,nm thick film ($T_c = 8.5$\,K, $R_{\square}=250\,\Omega/\square$), we obtained $\xi \approx 11$\,nm, $\lambda_\text{L} \approx 502$\,nm, and $\Lambda \approx 56$\,µm. 
This result suggests that the micro-bridge can be considered as a narrow wire ($w\ll\Lambda$). We used these values for the simulations of Section \ref{sec: 2sup} and \ref{sec: 3sup}. In Section \ref{sec: 5sup}, we estimated the value of $\lambda_\text{L}$ from the experiment.

\newpage

\section{TDGL simulation of the vortex diode effect}\label{sec: 2sup}

\begin{figure*}[htbp!]
\includegraphics[width=16cm]{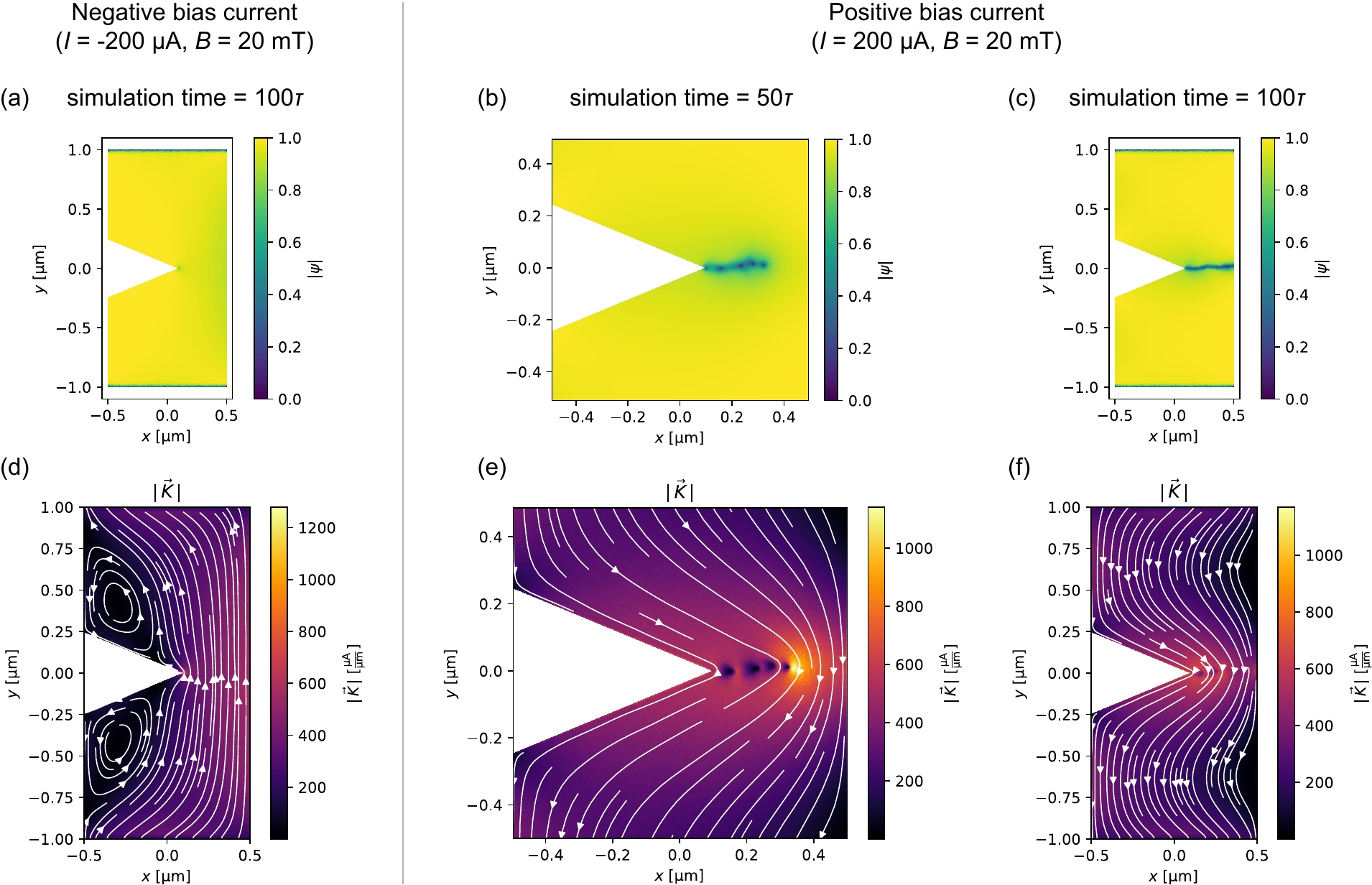}
            \caption{TDGL simulation of the diode effect. The figures on the top illustrate the order parameter when a 20\,mT field is applied with (a) $I=-200$\,µA after a simulation time of $100\tau$; (b) $I=200$\,µA at $50\tau$ (close-up fo the constriction); and (c) $I=-200$\,µA at $100\tau$. The bottom figures (d), (e), and (f) show the sheet current distribution for the same three different conditions. Parameters used in pyTDGL: $\xi = 10$\,nm, $\lambda_L = 500$\,nm, $d = 15$\,nm.}\label{fig: 1sup}
        \end{figure*}

Solving the time-dependent Ginzburg-Landau (TDGL) equations for the used device geometry helps in understanding the physical mechanism behind the observed vortex diode effect. We used the PyTDGL Python package \cite{horn_pytdgl_2023} to perform TDGL simulations of our devices.
Figure \ref{fig: 1sup} shows the results for a diode with a 600\,nm notch width, and a $45^{\circ}$ notch angle ($\xi = 10$\,nm, $\lambda_L = 500$\,nm, $d = 15$\,nm).
We observed the diode effect by applying an out-of-plane magnetic field of 20\,mT (entering the surface) and biasing the device with a positive or negative 200\,µA current (positive from the top to the bottom).
With the negative current, no vortices penetrated the film after a simulation time of $100\tau$ ($\tau=\mu_0\sigma\lambda^2_\text{L}$) and the device remained in the superconducting state. With the positive current, after $50\tau$, four vortices entered into the film from the notch tip, which is the region with the highest current density (due to current crowding) and the lowest surface barrier. At a $100\tau$ simulation time, the wire became normal at any point across the constriction width. This result confirmed that the critical current for which vortices enter the wire depends on the bias orientation. 

The value of applied field (20\,mT) was not chosen to maximize the efficiency but to demonstrate the device operation principle in short simulation times. 
A more accurate study would consist of sweeping the bias current and magnetic field to obtain I-V curves of the diode and reproduce the field dependence of $\eta$ (similar to Figure 2b).
In simulation, the field for peak efficiency, which increases with the critical current density, is higher than the value in Figure 2b (3.3\,mT) because the material is modeled as a perfect crystal without defects. A defect on the right edge of the wire decreases the field for peak efficiency.
Moreover, this model is incomplete because it does not include the Joule heating contribution to the hotspot generation and growth.

\section{Rectification efficiency for different diode geometries}\label{sec: 3sup}
   
 \begin{figure*}[htbp!]
          
            \includegraphics[width=13.3cm]{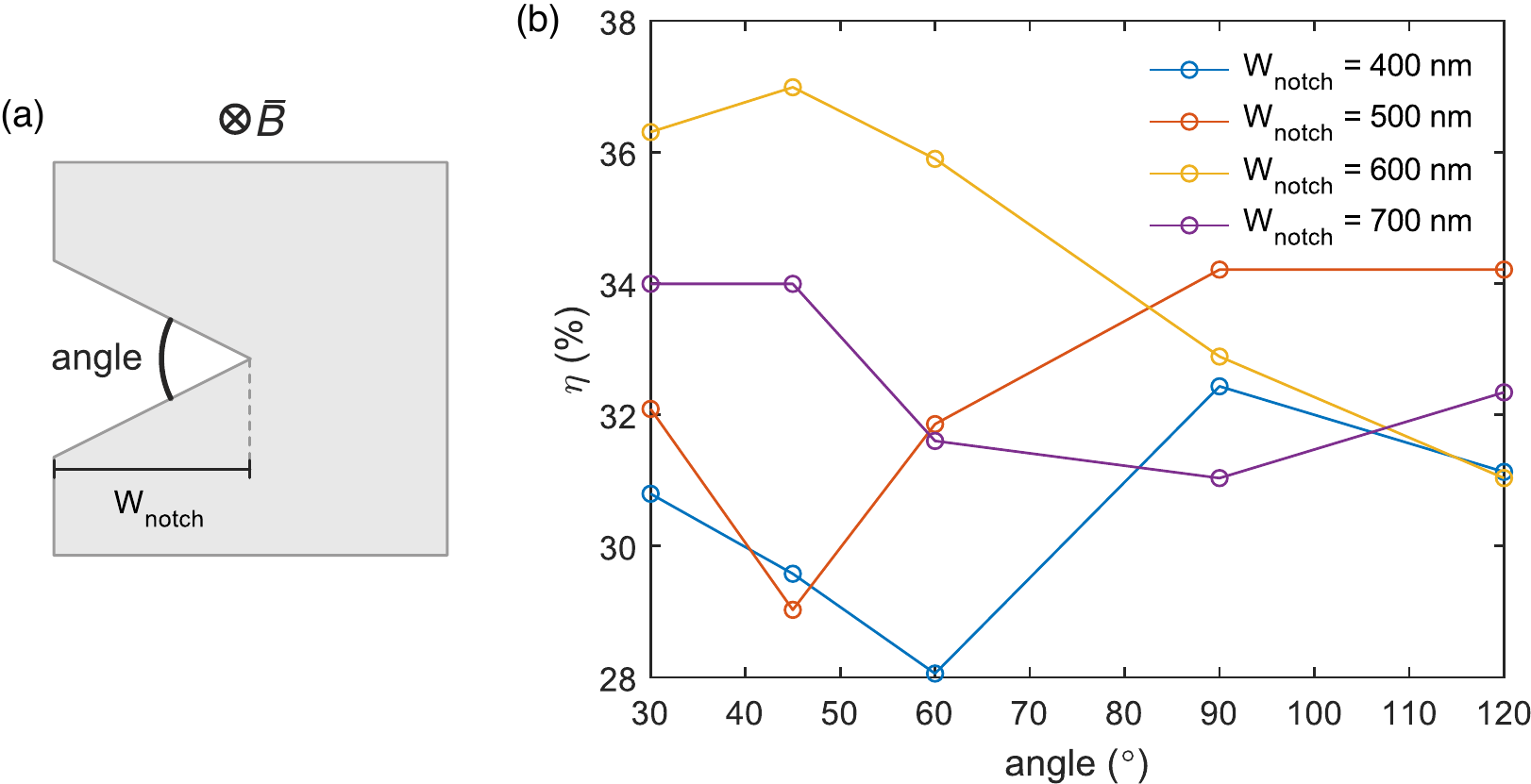}
            \caption{Experimental rectification efficiency for different diode geometries. (a) illustration of the diode geometry showing relevant parameters. (b) Rectification efficiency as a function of the notch angle for different notch width $W_\text{notch}$ ($d=14$\,nm). Each data point was obtained by extracting critical currents averaging over 1000 samples and sweeping the magnetic field to maximize $\eta$. The relative standard deviation $\sigma_{\eta}/\eta$ was lower than 2\,\% for every point.}\label{fig: 2sup}
        \end{figure*}

We fabricated and characterized 20 diodes with different notch widths ($W_\text{notch}$) and angles (the geometry is shown in Figure \ref{fig: 2sup}a) on the 14\,nm thick film to study how the rectification efficiency $\eta$ depends on the device geometry.
Figure \ref{fig: 2sup}b shows $\eta$ as a function of the angle for different notch widths. Each data point was found by sweeping the applied magnetic field to maximize the efficiency of a single device.  
$\eta$ varied from 28\,\% to 37\,\%, and the highest value was achieved with a 600\,nm notch width and a $45^{\circ}$ angle. 
Ideally, as observed in simulations of wires much larger than the notch from Benfenati et al. \cite{benfenati_vortex_2020}, a sharper angle should correspond to a lower left vortex barrier, and intuitively to a higher $\eta$. With a notch width similar to the wire width, we did not clearly see this dependence for every $W_\text{notch}$. Only for $W_{notch} = 600$\,nm, $\eta$ approximately followed the expectations.
Therefore, we investigated this effect more by analyzing how angle and notch width alter the sheet current density ($K$) distribution across the constriction (from the notch tip to the right wire edge) in TDGL simulations.

\begin{figure*}[htbp!]
      
        \includegraphics[width=16cm]{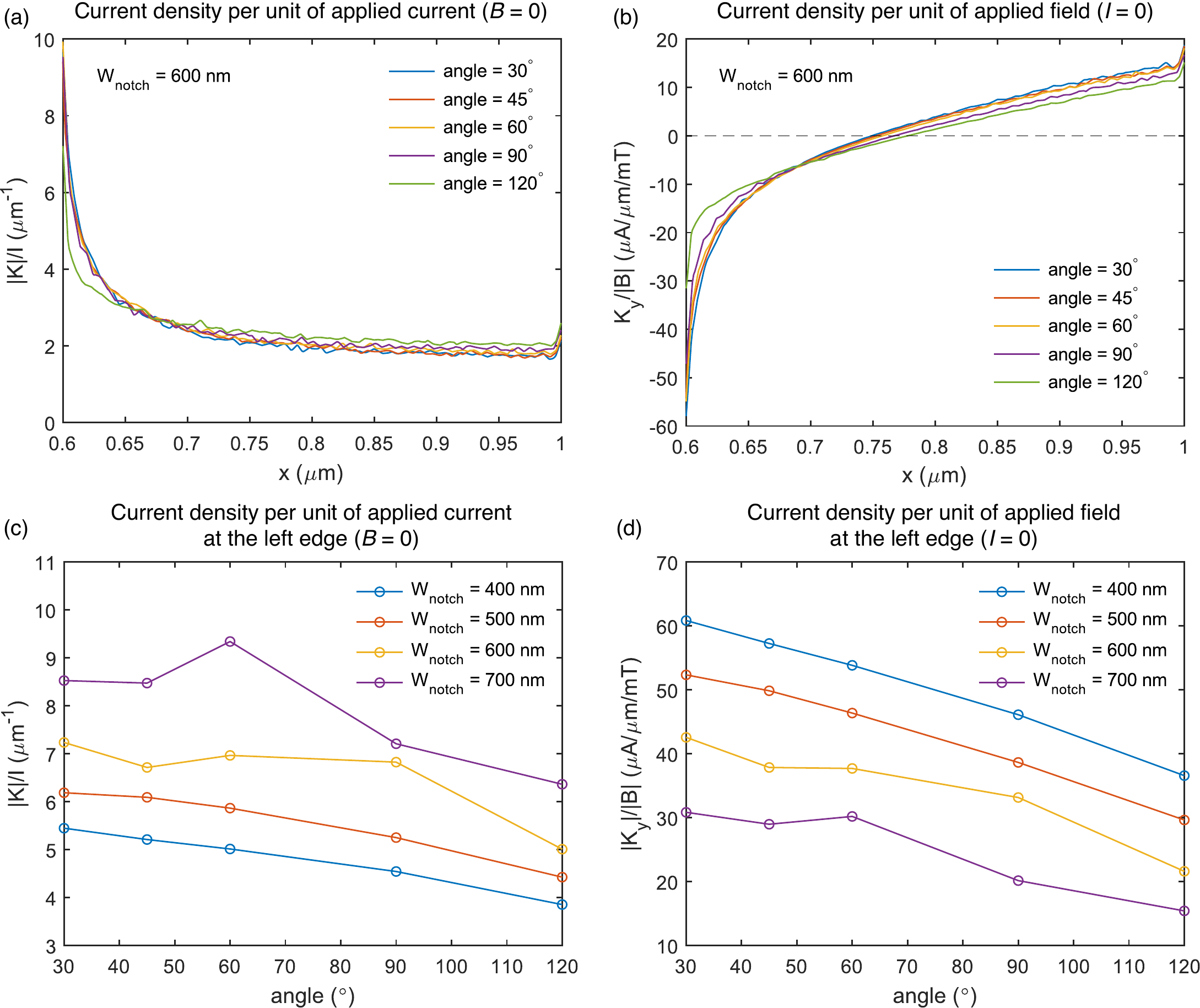}
        \caption{Distributions of the applied current density and the Meissner current density for different diode geometries in TDGL simulations. (a) Sheet current density per unit of applied current across the constriction (from the notch tip to the right edge of the wire), for different notch angles, with $W_\text{notch}= 600$\,nm and $B=0$\,mT. (b) Sheet Meissner current density per unit of applied field across the constriction, for different notch angles, with $W_\text{notch}= 600$\,nm and $I=0$\,A. (c) Sheet current density at the notch tip per unit of applied current as a function of notch angle, for different $W_\text{notch}$, with $B=0$\,mT. (c) Sheet Meissner current density at the notch tip per unit of applied field as a function of notch angle, for different $W_\text{notch}$, with $I=0$\,A. The values at the notch tip were obtained averaging between $x = W_\text{notch}$ and $x = W_\text{notch}+\xi$. Parameters used in pyTDGL: $\xi = 10$\,nm, $\lambda_L = 500$\,nm, $d = 15$\,nm.}\label{fig: 3sup}
    \end{figure*}

The current density in proximity to the notch tip (which defines the critical current for fields below the value for peak efficiency) depends on the combination of the bias current and the Meissner current. The notch geometry, through the current crowding effect, influences the current distribution for both bias and Meissner currents.
We analyzed the effects of the geometry on bias and Meissner currents separately. Figure \ref{fig: 3sup}a shows the current distribution when a bias current is applied with zero magnetic fields, for different angles. 
Figure \ref{fig: 3sup}b shows the distribution of the Meissner current with zero applied current, for different angles.
These plots show that the distributions are very similar to each other even if changing the angle.

We further analyzed the problem by calculating the current density $|K|$ at the left edge, which defines the critical current, for different notch widths and angles. In particular, the slope of the $|I^-_\text{c}|$ curve versus field in Figure 2b depends on the Meissner current density per unit of applied field at the left edge (more 
 details are in Section \ref{sec: 5sup})
 We calculated $|K|$ at the edge by averaging on the curves of Figures \ref{fig: 3sup}a and \ref{fig: 3sup}b between $x = W_\text{notch}$ and $x = W_\text{notch}+\xi$. The averaging was used to take into account that the experimental sharpness of the tip might have differed from the ideal case in simulation. 

From Figure \ref{fig: 3sup}c and \ref{fig: 3sup}d, for a given angle, the notch width had an opposite effect on the bias current distribution and the Meissner current: with larger notches, the bias current density increased, while the Meissner current decreased. This could explain why in Figure \ref{fig: 2sup}b, there is no clear dependence of $\eta$ on the wire width. 
For both figures, at $W_\text{notch} = 400$\,nm, 500\,nm  decreasing the angle clearly increased $|K|$, while at $W_\text{notch} = 600$\,nm, 700\,nm the dependence was less defined. Therefore, we could not exactly map the simulations to the experimental results. However, Another important factor in defining the positive and negative critical current is the experimental edge roughness at the right and left edges, which could alter the vortex barriers, and thus should be studied in simulations. In particular, the presence of a defect on the right edge would lower the right critical current density and thus the peak efficiency (the peak efficiency is reached when right and left current density are equal). 
In the future, a more accurate simulation that combines the effects of bias current, magnetic field, and edge roughness, could give more insights into this effect. 
   
\section{Retrapping current of the diode}
   
\begin{figure*}[htbp!]
          
            \includegraphics[width=8cm]{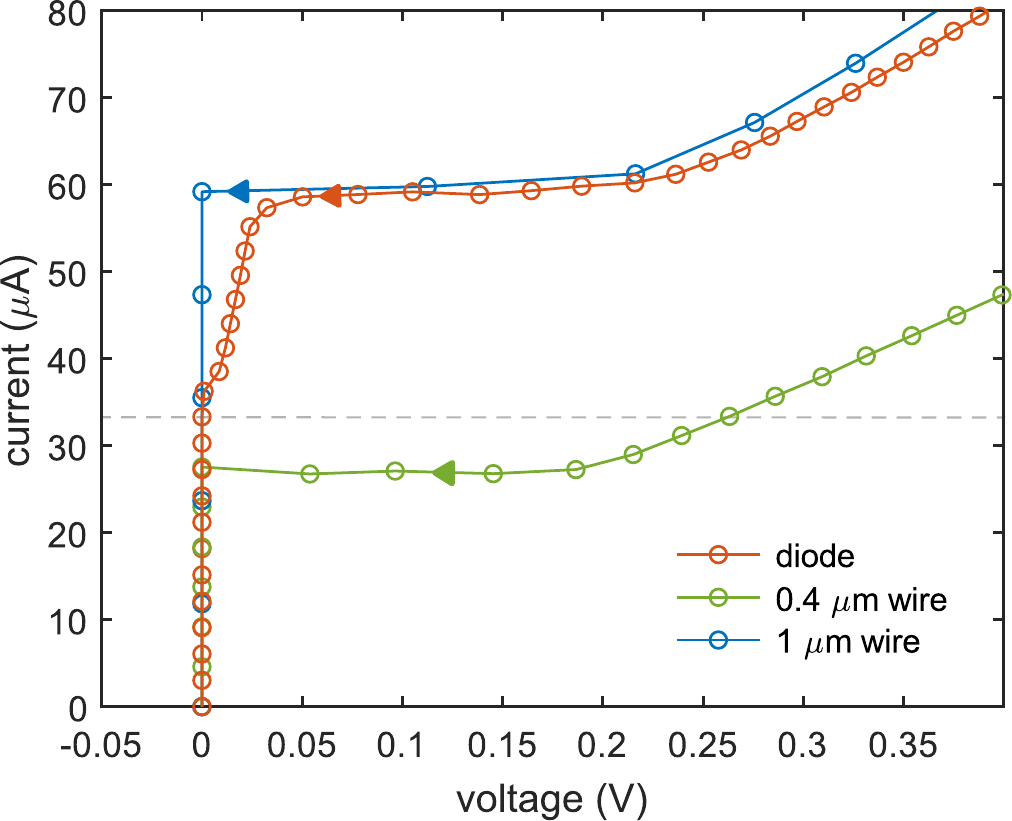}
            \caption{Retrapping current for different micro-bridge geometries. The figure shows the downward sweeps of I-V curves for a 1\,µm and 400\,nm wide wire without notch, and a diode with $45^{\circ}$ angle and $W_\text{notch} =400$\,nm. All the devices were fabricated on the 13\,nm thick film. The arrows indicate the sweep direction. The dashed line indicates the retrapping current of the diode $I_{r2}= (33\pm3)$\,µA for which the voltage drop becomes zero (at 36\,µA there still is a small voltage drop)}. \label{fig: 4sup}
        \end{figure*}

Figure \ref{fig: 4sup} shows the comparison between retrapping currents of a diode ($45^{\circ}$ angle, $W_\text{notch} =400$\,nm) and straight wires with width $w=400$\,nm and $w=1$\,µm. The high retrapping current of the diode $I_\text{r1}=(60\pm3)$\,µA is approximately equal to the value for $w=1$\,µm. Indeed, it corresponds to the hotspot retrapping in the 1\,µm wide part of the diode.
Below 50\,mV of voltage drop, the hotspot retraps in the region with $w<1$\,µm, and then completely disappears when it reaches the constriction at $I_\text{r1}=(33\pm3)$\,µA.
$I_\text{r1}$ is slightly higher than 28\,µA, retrapping current for $w=400$\,nm. The reason might be a small difference between the constriction width and the straight wire width.
   
\section{Field dependence of the critical current (d\,=\,9\,nm)}
   
\begin{figure*}[htbp!]
          
            \includegraphics[width=7.7cm]{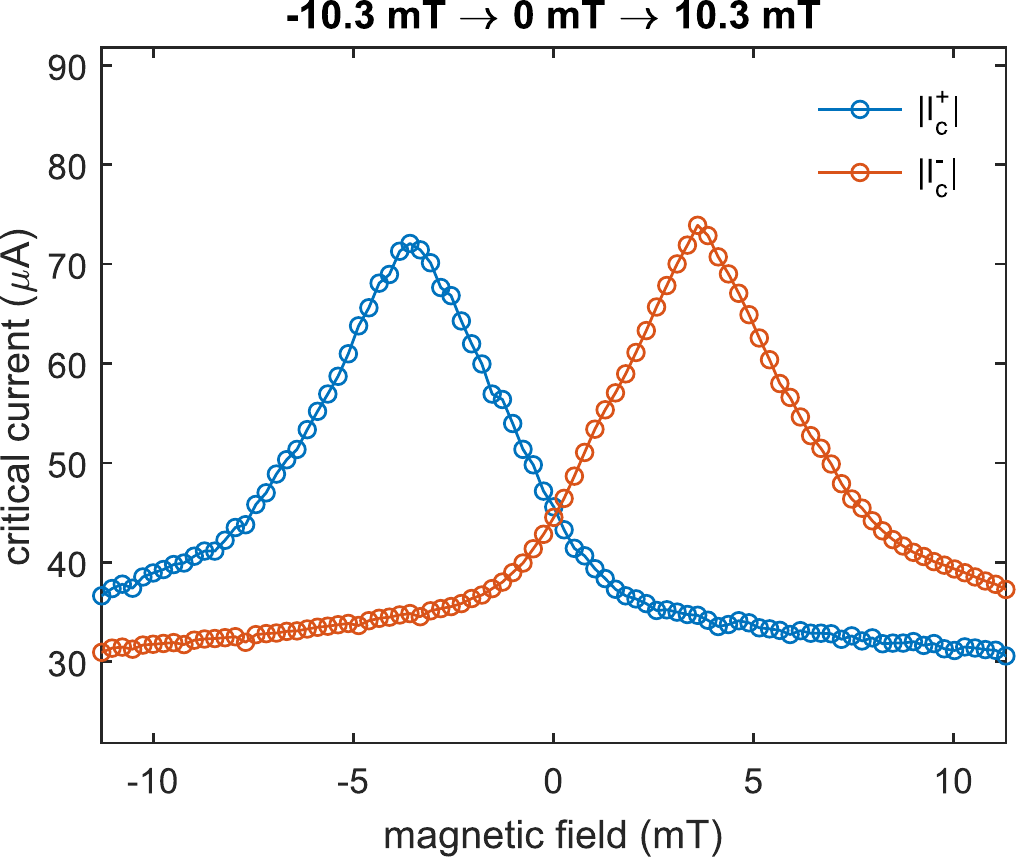}
            \caption{$|I^{+}_\text{c}|$ and $|I^{-}_\text{c}|$ as a function of the magnetic field (field swept from -10.3\,mT to 10.3\,mT.) for $d = 9$\,nm.} \label{fig: 5sup}
        \end{figure*}

Figure \ref{fig: 5sup} shows the positive and critical currents as a function of the magnetic field for the 9\,nm thick film. The asymmetry in the peak critical currents for $|I^-_\text{c}|$ and $|I^+_\text{c}|$ is 2.5\,\% of the average between the two peak currents. 

\section{Extraction of the penetration depth}\label{sec: 5sup}
   
\begin{figure*}[htbp!]
      
        \includegraphics[width=16cm]{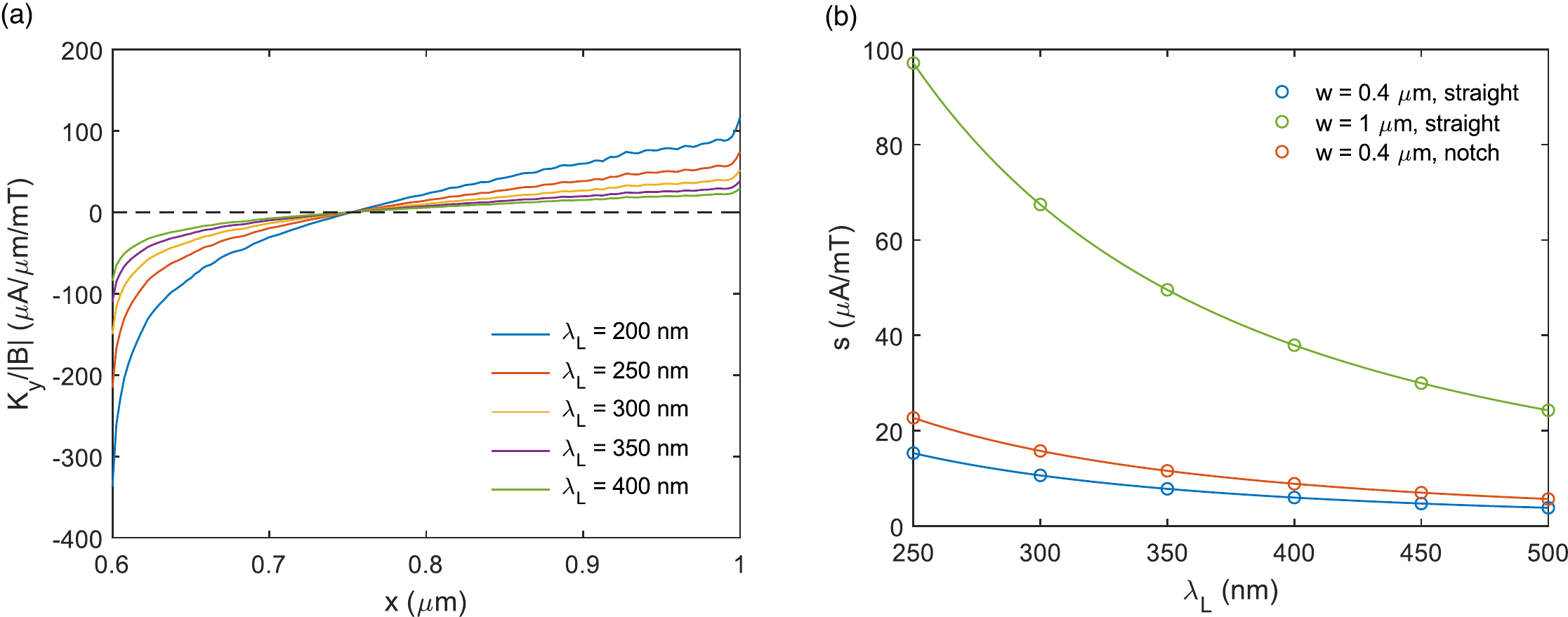}
        \caption{Dependence of the current density distribution on the London penetration depth in TDGL simulations. (a) Sheet Meissner current density per unit of applied field across the constriction, for different penetration depths, with $W_\text{notch}= 600$\,nm and $I=0$\,A. (b) Estimated slope $s=|K_B/B|_l/|K_I/I|_l$ of the $|I_c^-(B)|$ curve, as a function of $\lambda_\text{L}$ for straight wires without notch and a wire with notch. $w$ is the width of the constriction. Parameters used in pyTDGL: $\xi = 10$\,nm, $d = 15$\,nm.} \label{fig: 6sup}
    \end{figure*}

According to the theory for narrow and straight wires without defects ($w\ll\Lambda$), the slope of $|I^-_\text{c}|$ versus field should be $s \approx dw^2/2\mu_0\lambda_\text{L}^2= w^2/\mu_0\Lambda$ \cite{kuit_vortex_2008, plourde_influence_2001, vodolazov_superconducting_2005}. 
However, for our geometry with a defect width comparable to the wire width, the current distribution across the construction varies from the ideal and, therefore, the equation for the slope might be different. 
We compared the experimental slope of Figure 2b with the information from the TDGL simulations to estimate a correction factor for our geometry and extract $\lambda_\text{L}$ and Pearl length $\Lambda$. 

For a geometry with the notch, the current crowding effect influences the slope, since the Meissner current is maximum in proximity to the notch tip.
The critical current $I_\text{c}(B=0)$ without the magnetic field for a wire with no notches is $I_\text{c}(0)=J_\text{c} dw$. Assuming that $J_\text{c}$ does not change with the geometry, with current crowding due to the notch, the current density $J(x=0.6\,\text{µm}) = J_l$ at the left edge defines the critical current so that $I_c(0)=J_\text{c} d/|K_I/I|_l$, where $|K_I/I|_l$ is the sheet current density per unit of applied current at the left edge shown in Figure \ref{fig: 3sup}a (without notch $|K_I/I|_l = 1/w$). 
With an applied magnetic field (positive entering the surface), Meissner and applied current density add up at the left edge, modifying the critical current. 
The resulting current density on the left is $J_l = I|K_I/I|_l/d - |K_B/B|_lB/d$, where $|K_B/B|_l$ is the sheet current density per unit of field at the left edge (from Figure \ref{fig: 3sup}b). The negative critical current is reached when $J_l = J_\text{c}$ so we can define a formula for $|I^-_\text{c}|$ when the field is lower than 3.3\,mT (field for peak efficiency in Figure 2b):
\begin{equation}
    |I^-_\text{c}(B)| \approx \frac{J_\text{c}d}{|K_I/I|_l} + \frac{|K_B/B|_l}{|K_I/I|_l}B = I_\text{c}(0)+sB
\end{equation}
where $|K_B/B|_l$ is a function of $\lambda_\text{L}$.

Figure \ref{fig: 6sup}b shows how $s$ varies with $\lambda_\text{L}$ in simulation without the notch for $w=1\,\text{µm}$ and $w=0.4\,\text{µm}$, and with the notch ($45^{\circ}$ angle). Without the notch $s \approx dw^2/2\mu_0\lambda_\text{L}^2= w^2/\mu_0\Lambda$ following the theory for straight narrow wires ($w\ll\Lambda$) , and with the notch $s=1.48\,dw^2/2\mu_0\lambda_\text{L}^2=1.48\,w^2/\mu_0\Lambda$.
From the experiment on the 14\,nm thick diode, we obtained $s=22.5$\,µA/mT, which corresponds to $\lambda_\text{L} = \sqrt{1.48\,dw^2/2\mu_0s}\approx 250$\,nm and $\Lambda\approx 8.9$\,µm.
For the 9\,nm thick film, we measured $s=8$\,µA/mT, which corresponds to $\lambda_\text{L}\approx 330$\,nm and $\Lambda\approx 24$\,µm.
Considering that the simulated and experimental tip sharpness might be different and that our assumption of constant $J_\text{c}$ might be wrong, the correction factor could be an overestimation. Therefore, we can conclude that $205\,\text{nm}\le\lambda_\text{L}\le250\,\text{nm}$, where 205\,nm was calculated without correction ($6\,\text{µm}\le\Lambda\le8.9\,\text{µm}$). For $d=9$\,nm, $270\,\text{nm}\le\lambda_\text{L}\le330\,\text{nm}$ and $16\,\text{µm}\le\Lambda\le24\,\text{µm}$. These values are in agreement with results in the literature \cite{luo_niobium_2023, medeiros2022investigation}
   
\section{Extraction of the critical field}

The critical field $B_\text{s}$ for the penetration of vortices is the value for which the surface barriers are completely suppressed and the critical current becomes zero \cite{shmidt_critical_1970}.
For a narrow and straight wire ($w\ll\Lambda$) it is calculated using $B_s = \phi_0/2\pi\xi w$ \cite{maksimova_mixed_1998} or $B_s \approx \mu_0 2J_\text{c}\lambda^2_\text{L}/w$ \cite{plourde_influence_2001}.
We extracted the value of $B_\text{s}$ by finding the intersections of the linear fits with the x-axis in Figure 2b \cite{vodolazov_superconducting_2005}. The values were (-5.5, +10.6)\,mT for the $|I^-_c(B)|$ curve and (-12, +5.7)\,mT for $|I^+_c(B)|$, with $d = 14$\,nm. For $d = 9$\,nm, the four fields were (-5.4, +13.4)\,mT and (-13.6, +5.6)\,mT. 
The lower value (around 5.6\,mT for $d = 14$\,nm) should depend on the left edge while the high value (around 11.3\,mT for $d = 14$\,nm) on the part of the right edge with the highest current density. Since we did not include in the design a defect on the right edge, we do not know which is the weak point defining the peak of efficiency, the high critical field, and the slope of $I_\text{c}(B)$ above 3.3\,mT. Considering that the slope above 3.3\,mT was comparable to the one below 3.3\,mT, the weak point on the right edge should have been in proximity to the constriction. 

The reduction in $B_\text{s}$ due to the presence of the notch has been already calculated for a similar geometry as a function of the angle: $b_\text{s} = (2\xi/\Lambda)^{1-\pi/(2\pi-\theta)}B_\text{s}$, where $\theta$ is the angle ($\pi/4$) \cite{buzdin_electromagnetic_1998}. This equation might differ for our device because the notch width is comparable to the wire width, however, we used it as a rough estimation, obtaining $b_\text{s}\approx 5.7$\,mT for $w=400\,\text{nm}$ and $\xi=10.8$\,nm. The value is in close agreement with the experimental values (5.6\,mT). Using a more realistic value of $\xi\approx5$\,nm \cite{luo_niobium_2023}, the value is 8.9\,mT, which is still comparable to our experiment.

For a straight wire without defects, we estimated a critical field of $B_\text{s} = \phi_0/2\pi\xi w\approx 30$\,mT with $w=1$\,µm and $B_\text{s} \approx 75$\,mT with $w=0.4$\,µm, These values are on the same order of magnitude as the experimental field limited by the right edge (around 11.3\,mT). The difference might come from the presence of a defect on the right edge. 

From Figure 2b, we could also extract the critical current density $J_\text{c}$ of the film by looking at the critical current for zero field $I_\text{c}(0)=124.5$\,µA and using equation 1, we obtained $J_\text{c} \approx I_\text{c}(0)|K_I/I|_l/d \approx 6.0 \times10^{10}$\,A/m$^2$. 
Assuming that $J_\text{c}$ does not change with the edge geometry, with this value, we could estimate the critical field for a straight wire without defects using the second equation: $B_s \approx \mu_0 2J_\text{c}\lambda^2_\text{L}/w\approx 9.4$\,mT for $w=1$\,µm, and 23\,mT for $w=0.4$\,µm.

%ratio
For $d=14$\,nm, the experimental ratio $h_\text{s}/B_\text{s}$ was 2.02 (using the average values between the two critical current curves) and for $d=9$\,nm it was 2.46. Therefore the thinner film had a 1.22 times higher asymmetry. This result confirms the reduction of the critical field and the asymmetry of the surface barriers increases with the thickness as predicted by the theory \cite{vodolazov_superconducting_2005, plourde_influence_2001}. 
Considering the ideal case of a maximum field reduction $(2\xi/\Lambda)^{3/7}$ \cite{buzdin_electromagnetic_1998}, and using the values of $\lambda_\text{L}$ extracted for the two films, $b_\text{s}/B_\text{s}$ should scale as $(d/\lambda_\text{L}^2)^{3/7}$: $((14/9)(330/250)^2)^{3/7}=1.53$ instead of 1.22.
This difference suggests that the field reduction was not optimal, probably due to not ideal notch sharpness or the presence of defects on the right edge, which lowered $B_\text{s}$. The reduction seemed to scale as $(d/\lambda_\text{L}^2)^{1/5}$ (using only the data from these two films).
Another explanation might be that the difference in the estimated $\lambda_\text{L}$ between the two films came from a different notch sharpness rather than film properties. In this case, the two penetration depths would be equal and the ratio would scale as $d^{3/7}$: $(14/9)^{3/7}=1.21$, similar to the experimental value. 
   
\newpage

\section{Low-frequency half-wave rectification}

        \begin{figure*}[htbp!]
          
            \includegraphics[width=13.3cm]{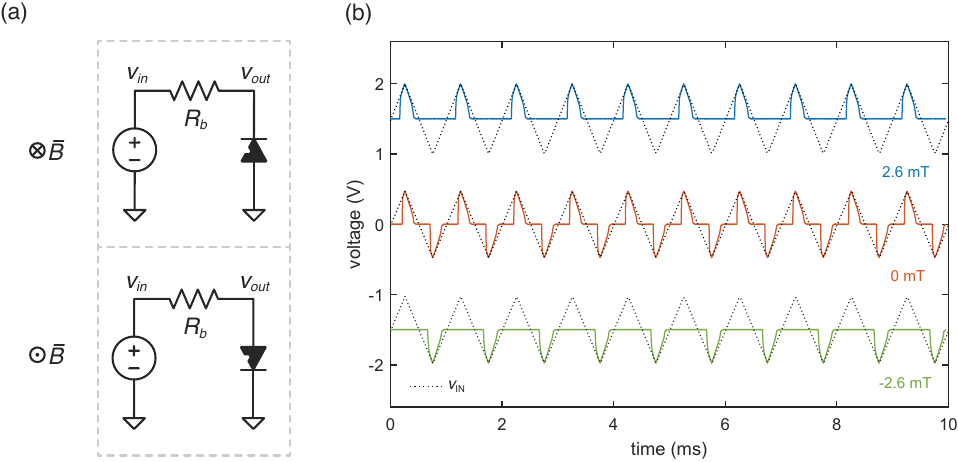}
            \caption{Half-wave rectification of a triangular wave at low frequency. (a) Circuit schematic of a current-biased superconducting diode for opposite diode polarities. This setup was used both to measure I-V curves and to demonstrate half-wave rectification ($R_\text{b} = 10$\,k$\Omega$). (b) Half-wave rectification of a 1\,kHz triangular wave for zero, positive, and negative magnetic fields. The triangular waves in dashed lines ($v_\text{IN}$) are theoretical input signals scaled to have the same amplitude of $v_\text{OUT}$. Traces are vertically shifted for clarity. Each trace is the average of 100 different traces acquired in sequence, to increase the signal-to-noise-ratio.} \label{fig: 7sup}
        \end{figure*}
        
        %rectification diode
        A typical use for a single superconducting diode is half-wave rectification of a periodic signal.
        Using the same setup as in the I-V curve measurement (a simplified schematic is shown in Figure \ref{fig: 7sup}a), we confirmed half-wave rectification of a triangular wave at 1\,kHz, shown in Figure \ref{fig: 7sup}b. The maximum current was set to a value in between the critical current of the wire with zero field, and $I^{-}_\text{c}$ ($I^{+}_\text{c}$ for negative magnetic field). With zero applied field, a resistive state was generated on both the positive and negative amplitudes of the curve. For positive magnetic fields, only positive currents generated a voltage on the output. An analogous and opposite effect was observed for negative fields.

\newpage

\section{High-frequency half-wave rectification}
   
\begin{figure*}[htbp!]
          
            \includegraphics[width=10cm]{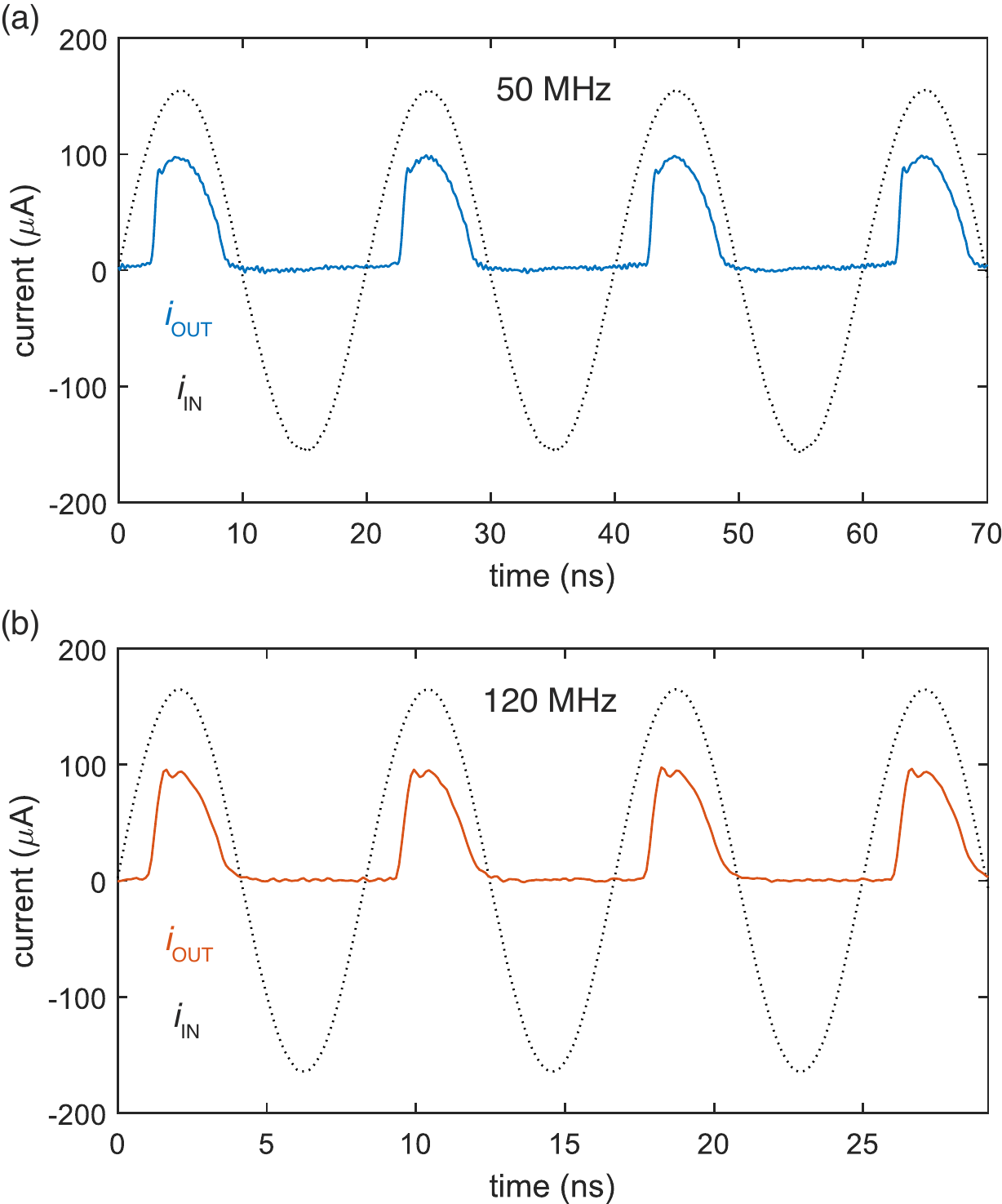}
            \caption{Half-wave rectification of a sinusoidal signal at (a) 50\,MHz and (b) 120\,MHz. For both plots, the dashed line is the input signal $i_\text{IN}$ and the full line is the output signal $i_\text{OUT}$, which corresponds to the current flowing through the load resistor $R_L$. The signals are averaged over 100 waveforms. For figure (b), a sinusoidal wave at 120\,MHz, with amplitude 3.4\,µA, was subtracted from $i_\text{OUT}$ to remove the effect of the parasitic inductance of wire bonds. These parasitics decreased the applied input current by only 0.5\,\% at 120\,MHz, according to SPICE simulations. } \label{fig: 8sup}
        \end{figure*}

We demonstrated half-wave rectification of sinusoidal signals up to 120\,MHz using the differential setup shown in Figure \ref{fig: 9sup}. We did not test the diode at higher frequencies due to the limits of the AWG.
Figure \ref{fig: 8sup} shows input and output waveforms at 50\,MHz and 120\,MHz. 
The ratio between peak output and input currents is 0.63 at 50\,MHz, and 0.54 at 120\,MHz.

\section{Measurement setup for high-frequency half-wave rectification}
   
\begin{figure*}[htbp!]
  
            \includegraphics[width=13.3cm]{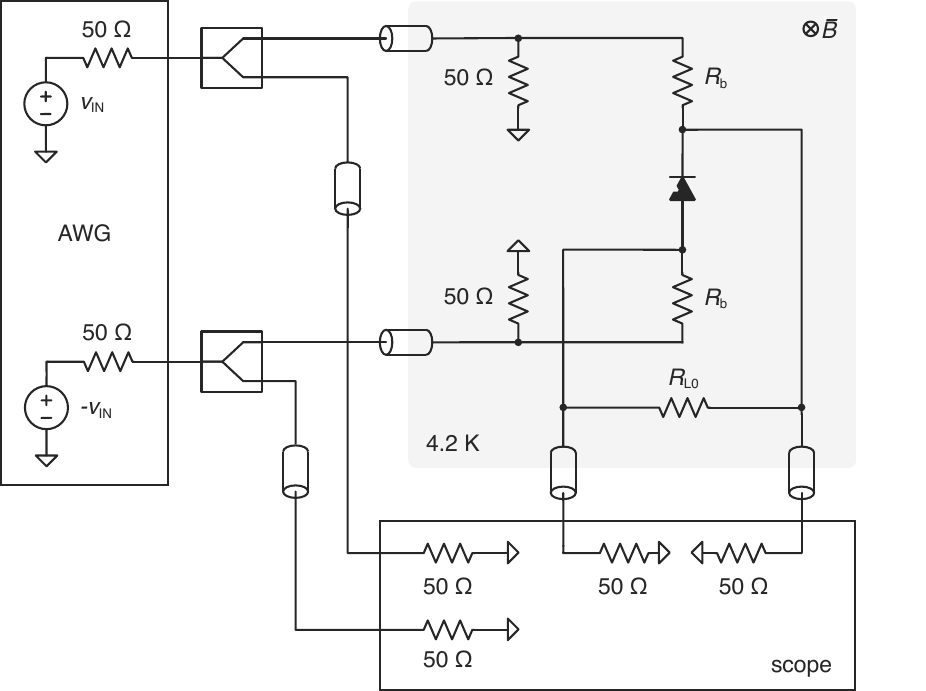}
            \caption{Circuit schematic of the measurement setup for high-frequency half-wave rectification. An AWG generates two identical but inverted signals ($v_\text{IN}$ and $-v_\text{IN}$) that are sent to two splitters. For each splitter, one port is connected to the scope, and the other port is connected to a bias surface-mount resistor $R_\text{b}$ and a parallel surface-mount 50\,$\Omega$ resistor (this resistor ensures impedance matching between AWG and device), to drive one of the two ports of the diode. 
            A surface-mount resistor $R_\text{L0}$, connects the two ports of the diode. The output voltage across $R_\text{L0}$ is differentially measured by the oscilloscope. 
            All the components in the gray box are at 4.2\,K. The magnetic field is applied to all the devices at low temperatures. Coaxial cables connect the circuit to room-temperature electronics.}\label{fig: 9sup}
        \end{figure*}

Figure \ref{fig: 9sup} shows the circuit schematic of the differential measurement setup used to characterize the rectification efficiency of single diodes for frequencies above 100\,kHz (described in the Methods section).
This setup was used also to demonstrate half-wave rectification at 50\,MHz and 120\,MHz.
 We used the oscilloscope to acquire both the input and output signals. The input current waveforms showed in \ref{fig: 8sup} were extracted by subtracting the two input signals ($v_\text{IN}$ and $-v_\text{IN}$) and dividing by $2R_\text{b}$. The output currents were extracted by subtracting the two output voltages and dividing them by the equivalent parallel resistance ($R_\text{L}$) of $R_\text{L0}$ and the 100\,$\Omega$ differential resistance of the scope. 
   
\section{Frequency limits of the diodes and temperature dependence of the rectification efficiency}\label{sec: 11sup}
   
\begin{figure*}[htbp!]
\includegraphics[width=16cm]{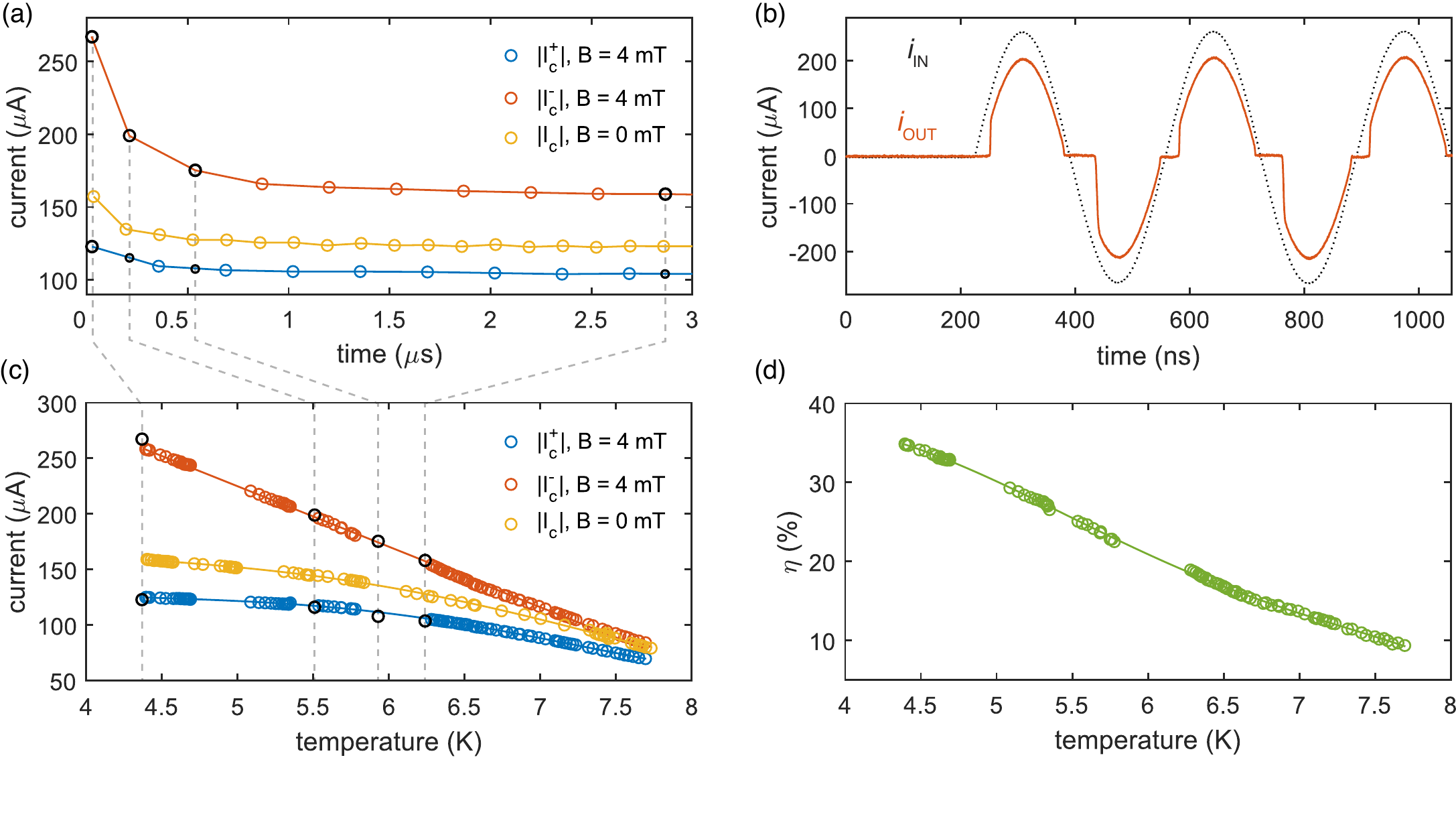}
            \caption{Experimental observation of a transient behavior for the rectification efficient in burst signals and relation to the temperature dependence of $\eta$. (a) Evolution of $|I_c^-|$, $|I_c^+|$, and $|I_c(B=0)|$ in time for a burst signal (30 cycles, 100\,µs burst period) at 3\,MHz. The currents and associated time were extracted by looking at positive and negative switches in the burst. The first point of the $|I_c^-|$ curve was obtained by inverting the polarity of the input signal so that the first switching event in the burst was related to $|I_c^-|$. (b) Input (dashed line) and output (full line) burst signals used to extract the critical currents (averaged over 100 traces). (c) $|I_c^-|$, $|I_c^+|$, and $|I_c(B=0)|$ as a function of temperature, with associated polynomial fits. The measurement was performed by sweeping up the temperature with the field on. (d) Rectifiction efficiency as a function of temperature, extracted from (c). The linear fit between 4.4\,K and 6.5\,K has a slope of -8.9\,\%/K.} \label{fig: 10sup}
        \end{figure*}

We investigated the causes of rectification efficiency drop at frequencies above 1\,MHz for diodes on the 13\,nm thick film, used for the bridge rectifier. First, we compared the experimental frequency limits of diodes to the theoretical value for rectification, by calculating the time vortices take to cross the constriction ($w=400$\,nm) with the formula given by Vodolazov et al. \cite{vodolazov_superconducting_2005} $t = 1.5 B_{c2} w \sigma_n/\langle j \rangle$, where $B_{c2} = \phi_0 / 2\pi \xi^2$ ($\xi=10$\,nm) is the second critical field, $\sigma_n=1/R_{\square} d$ ($R_{\square}=170$\,$\Omega/\square$, $d = 13$\,nm) is the normal conductivity of the material, and $\langle j \rangle \approx I_b/d w$ ($I_b = 100$\,µA) is the average current density in the wire. 
We obtained $t=47\,$ps, which corresponds to 21.3\,GHz, a much higher value than the frequency limit observed in Figure 2c ($\eta$ starts rolling off around 1\,MHz).
Another speed limitation might be related to the quasi-particle relaxation time $\tau_\text{E}$ of NbN but its value ranges from 12\,ps \cite{zhang_quasiparticle_2020} to 3.5\,ns \cite{cirillo_quasiparticle_2011} in literature. 
Therefore, we believe Joule heating, rather than vortex velocity or $\tau_E$ of the material, is mainly limiting the rectification efficiency of the diode above 1\,MHz.

We applied a 3\,MHz sinusoidal wave in burst mode (30 cycles) to a single diode (using the setup in Figure \ref{fig: 9sup}) to study how the efficiency evolves over time before reaching a steady state and thus understand if the transient is related to Joule heating.
Figure \ref{fig: 10sup}a shows the values of $|I^+_c|$, $|I^-_c|$, and $|I_c|$ (without field) for 10 burst cycles applied at 4.4\,K. Figure \ref{fig: 10sup}b shows 2.5 cycles of the burst signal used to extract the critical currents. 
The three critical currents decayed over 10 cycles. In particular, $|I^-_c|$ decreased by 42\,\% and $|I^+_c|$ decreased by 12\,\%, suggesting that the transient behavior was not caused by charging or discharging passive components in the measurement setup. In such a case, the relative current decrease would have been identical. To find a possible cause of this asymmetry in the decrease, we characterized the temperature dependence of the critical currents.
Figure \ref{fig: 10sup}c shows $|I^-_c|$, $|I^+_c|$, and $|I_c|$ versus temperature extracted at 1\,kHz (below the $\eta$ roll-off). $|I^-_c|$ had a linear dependence between 4.4\,K and 6.5\,K, while $|I^+_c|$ scaled sub-linearly (the associated value of $\eta$ versus temperature is shown in Figure \ref{fig: 10sup}d).  
We matched the paired values of $|I^+_c|$ and $|I^-_c|$ at 4 of the 10 cycles in Figure \ref{fig: 7sup}a to the ones in Figure \ref{fig: 7sup}c, finding correspondence with a maximum 3\,\% relative error on the currents. 
This result suggests that the power dissipated by the hotspot resistance at 3\,MHz caused a local heating from 4.4\,K to approximately 6.25\,K, and thus a decrease in efficiency from 35\,\% to 19\,\%. The efficiency at steady state (19\,\%) was different from the value obtained in Figure 2c (25\,\%) for the same frequency because the applied current amplitude $I$ used in this measurement was 1.25 times higher than the one used in Figure 2c (260\,µA instead of 208\,µA). Indeed, assuming the Joule heat $Q_J$ scales like $I^2$ and the temperature change $\Delta T$ scales linearly with $Q_J$, we can say that $\Delta\eta(I) \propto \Delta T \propto Q_J \propto I^2$, and therefore $\Delta\eta(260\,\text{µA})/\Delta\eta(208\,\text{µA})=1.25^2 = 1.56$. This ratio matches well the experimental ratio $(35\,\%-19\,\%)/(35\,\%-25\,\%) =1.6$, further suggesting that the Joule heating strongly influences the device performance and that the drop in efficiency is related to the amplitude and shape of the input signal. 

In this analysis, both $|I_c^+|$ and $|I_c^-|$ at steady state were lower than the values at time zero but Figure 2c shows the steady state value of $|I_c^+|$ slightly increased with frequency. This suggests overheating was negligible and other effects dominated the change in $|I_c^+|$ in the measurement for Figure 2c.
Joule heating was negligible because $|I_c^+|$ was extracted by sweeping up the signal amplitude until positive-voltage pulses were observed, and therefore no dissipation occurred before the pulses appeared. 
However, Joule heating was not negligible for $|I_c^-|$ in Figure 2c, because the current value was extracted by sweeping up the input until negative-voltage pulses appeared, while positive pulses were already dissipating heat (since $|I_c^-|>|I_c^+|$).
Possible causes of the $|I_c^+|$ increase might be related to the measurement setup but, from comparisons between simulation and experiment, we estimated that parasitic inductances of the wire bonds altered the input signal by only 0.5\,\% at 120\,MHz. Therefore, the increase might come from other setup parasitic or internal physical effects we did not consider. For example, increasing the frequency might increase the probability of having vortices pinned to defects in the wire while the hotspot is retrapping, and their presence might alter the surface barriers and thus the critical current \cite{plourde_influence_2001}.

        \begin{figure*}[htbp!]
\includegraphics[width=16cm]{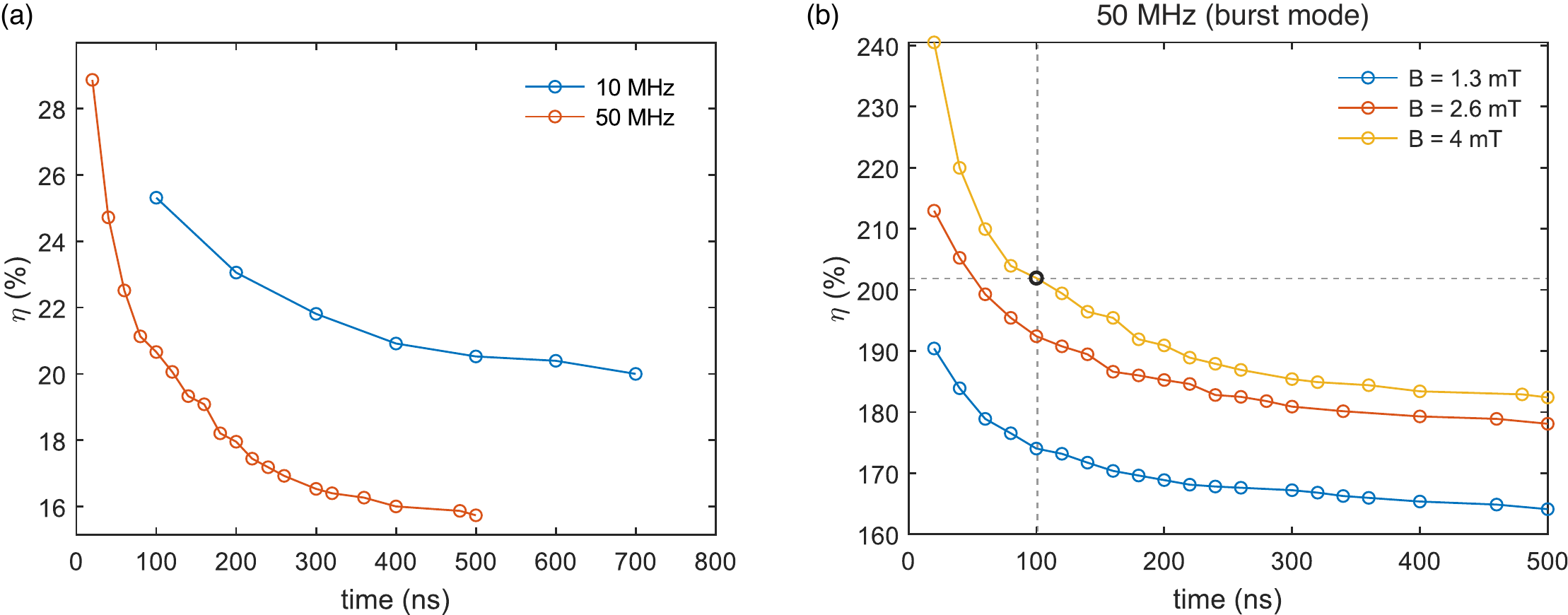}
            \caption{Experimental transient behavior of the rectification efficiency for different frequencies and applied fields. (a) $\eta$ as a function of time for a burst signal at 10\,MHz and 50\,MHz. At these frequencies, the method used in Figure \ref{fig: 10sup} was not accurate, so each point was obtained by sweeping the input amplitude until a stable pulse was observed in the associated cycle. (b) $|I_c^-|$ as a function of time for a 50\,MHz burst signal at different applied fields. The highlighted black point is associated with the maximum time the bridge rectifier could operate for at 50\,MHz. The value of the first cycle is already altered by the Joule heating of the first positive switch, we did not flip the polarity of the signal to extract $|I_c^-|$ at time zero, as in Figure \ref{fig: 10sup}.} \label{fig: 11sup}
        \end{figure*}

Since we observed a high drop of $|I_c^-|$ in time, which has a strong dependence on the temperature, we suggest the main contribution to the $\eta$ decrease with frequency comes from a local temperature rise for faster signals. Indeed, the heat dissipation to the substrate might be too slow to cool down the device at higher frequencies. 
Figure \ref{fig: 11sup}a shows $\eta$ as a function of time when a burst signal was sent at two different frequencies. As expected, the signal at higher frequency had a faster decay and lower efficiency at steady state.
Figure \ref{fig: 11sup}b shows $|I^-_c|$ versus time for a 50\,MHz sinusoidal signal in burst mode for different values of the applied magnetic field. Increasing the magnetic field, the current decay increased. $|I^-_c|$ dropped by 14\,\% for $B=1.3$\,mT, by 16\,\% for $B=2.6$\,mT, and by 24\,\% with $B=4$\,mT.
This result suggests that the field influenced the transient, and in particular, a larger field generated a larger decrease.  
We note that $|I^-_c|$ became 1.2 times lower than its initial value at 100\,ns (5 cycles), which approximately corresponds to the maximum time the rectifier could correctly operate for in Figure 4a (50\,MHz). 
Considering that the top limit of rectifier margins (in theory $|I^-_\text{c}|+I_\text{r}$) was around 260\,µA at low frequencies, we believe the circuit stopped rectifying after 100\,ns because $|I^-_\text{c}|+I_\text{r}$ became lower than the input current amplitude (240\,µA).

%%%%%%%%

\newpage
\section{Bridge rectifier with an asymmetric geometry}

\begin{figure*}[htbp!]
\includegraphics[width=8cm]{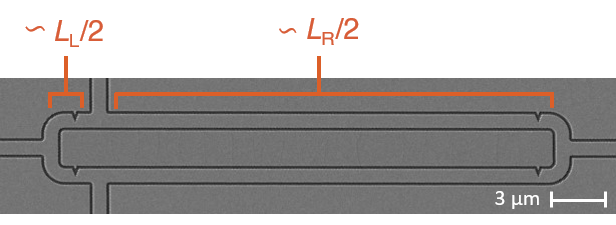}
            \caption{Scanning electron micrograph of the bridge rectifier with an asymmetric geometry. The red labels indicate the section of the wire in the superconducting loop associated with the kinetic inductances $L_\text{R}$ and $L_\text{L}$. The ratio $L_\text{R}/L_\text{L}$ is 10.} \label{fig: 12sup}
        \end{figure*}

Figure \ref{fig: 12sup} shows the scanning electron micrograph of the bridge rectifier we used to obtain most of the results of this work (the ratio $L_\text{R}/L_\text{L}$ is 10). 
We used the Electric Currents module of COMSOL Multiphysics to calculate the number of squares for each branch of the loop and choose the length of the right one to have a square ratio of 10. This design is independent of the value of sheet inductance.
The sheet inductance, used for the LTspice simulations shown in Sections \ref{sec: 14} and \ref{sec: 15}, was dominated by the kinetic inductance (about 30\,pH/$\square$).
Indeed, the geometric inductance of the loop is negligible for such geometry: we estimated a value of 30 pH by using the analytic formula for a rectangle loop with a wire diameter equal to the film thickness (worst case). This overestimation is much smaller than the total kinetic inductance of the loop (about 5\,nH).

\newpage

\section{Measurement setup for full-wave rectification}

\begin{figure*}[htbp!]
  
            \includegraphics[width=13.3cm]{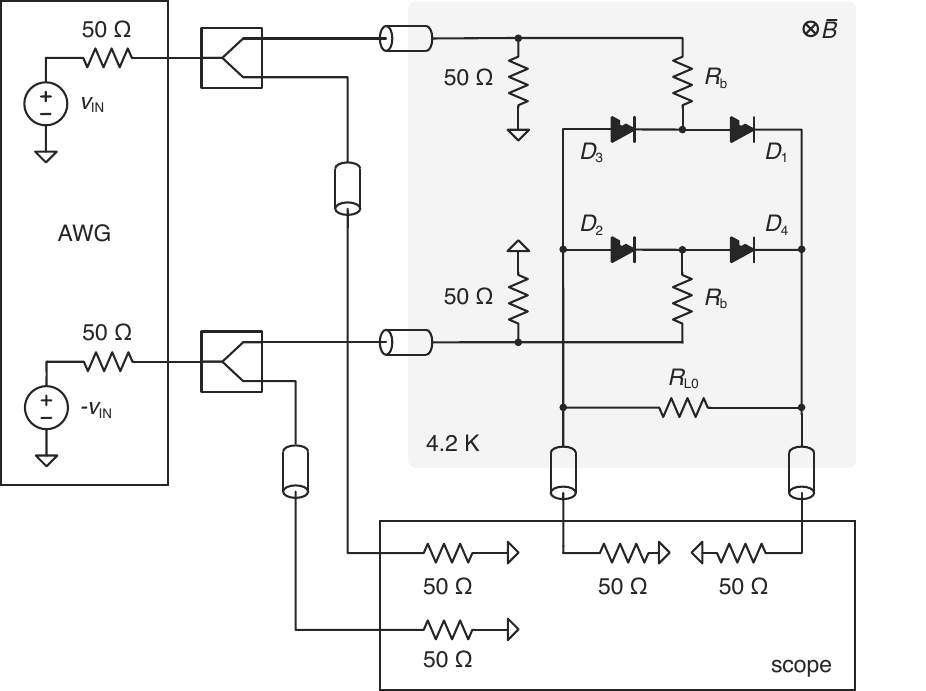}
            \caption{Circuit schematic of the measurement setup for full-wave rectification. An AWG generates two identical but inverted signals ($v_\text{IN}$ and $-v_\text{IN}$) that are sent to two splitters. For each splitter, one port is connected to the scope and the other port is connected to a bias surface-mount resistor $R_\text{b}$ and a parallel surface-mount 50\,$\Omega$ resistor (this resistor ensures impedance matching between AWG and device), to drive one of the two input ports of the rectifier. 
            A surface-mount resistor $R_\text{L0}$, connects the two output ports of the rectifier. The output voltage across $R_\text{L0}$ is differentially measured by the oscilloscope. 
            All the components in the gray box are at 4.2\,K. The magnetic field is applied to all the devices at low temperature. Coaxial cables connects the circuit to room-temperature electronics.}
            \label{fig: 13sup}
        \end{figure*}

 Figure \ref{fig: 13sup} shows the circuit schematic of the differential measurement setup described in the Methods Section (without amplifiers). 
 We used the oscilloscope to acquire both the input and output signals. The input current waveforms showed in the results were extracted by subtracting the two input signals ($v_\text{IN}$ and $-v_\text{IN}$) and dividing by $2R_\text{b}$. The output currents were extracted by subtracting the two output voltages and dividing them by the equivalent parallel resistance ($R_\text{L}$) of $R_\text{L0}$ and the 100\,$\Omega$ differential resistance of the scope.

\section{Full-wave rectification: experiment and simulation}\label{sec: 14}

\begin{figure*}[htbp!]
  
            \includegraphics[width=10cm]{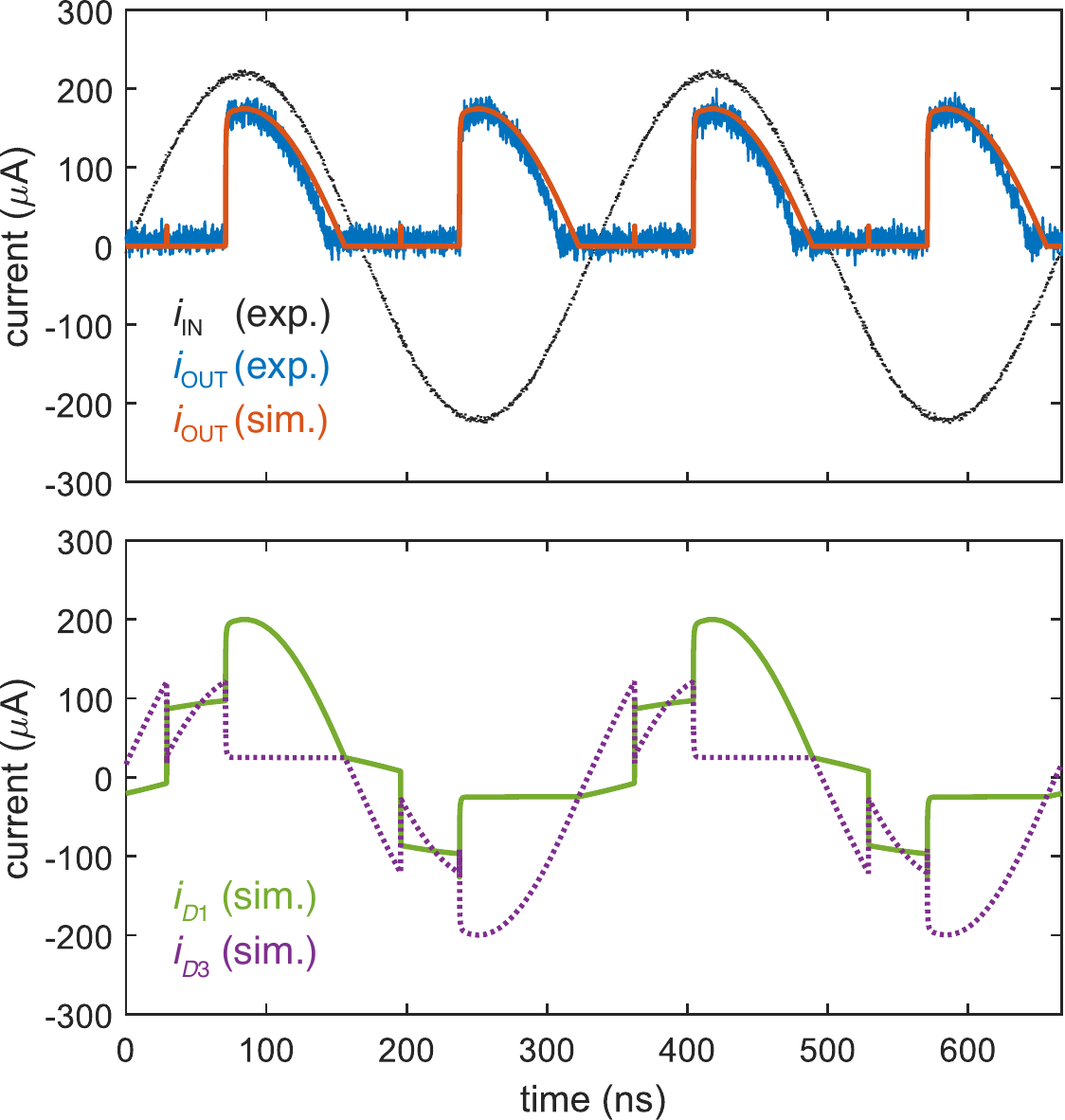}
            \caption{Comparison between experimental and simulated full-wave rectification at 3\,MHz. In the top panel, the black trace is the experimental input current of the rectifier, and the blue trace is the experimental output current without averaging. The red trace is the output current simulated with the parameters listed in the Methods section. The bottom panel shows the current through the diodes $D_1$ (full green trace) and $D_3$ (dashed purple trace) in the simulation. The current is positive when flowing from left to right in $D_1$ and from right to left in $D_3$ in Figure \ref{fig: 11sup}.}\label{fig: 14sup}
        \end{figure*}

Figure \ref{fig: 14sup} shows the comparison between the experimental and simulated full-wave rectification of a continuous sinusoidal signal at 3\,MHz. 
The top panel shows the input and output current of the rectifier, while the bottom panel shows the simulated current through the diodes $D_1$ and $D_3$.
The simulation was performed on the bridge rectifier with $L_\text{R}/L_\text{L} = 10$, using the LTspice model described in the Methods section. 

In the simulation, there was a short pulse of about 30\,µA at each semi-period before the large pulse (about 175\,µA). The small pulse was probably too fast to be observed in the experiment.
This behavior can be explained by looking at the bottom panel of Figure \ref{fig: 14sup}: when the positive input current is rising, first only $D_3$ switches (after about 35\,ns) because most of the current initially passes through the left branch; then $i_{D3}$ drops below the retrapping current $I_\text{r}=I_\text{r2}$ and a current $\approx i_\text{IN}-I_\text{r}$ lower than the critical current of $i_{D4}$ gets stored in the right branch; therefore, $D_3$ quickly resets and a short pulse is generated on the output; afterward, $i_{D3}$ keeps increasing until reaches the critical current again at around 70\,ns; at this point, the current diverted into the right branch is enough to make $D_4$ switch, keep the hotspot in $D_3$ stable, and generate the output pulse. 
To maintain a stable hotspot in both $D_3$ and $D_4$ (or $D_1$ and $D_2$), their current must be equal or higher than the retrapping current (or hotspot current), therefore the output signal is $i_\text{OUT} \approx i_\text{IN}-2I_\text{r}$ until $i_\text{IN}$ drops below $2I_\text{r}$ and the diodes reset. $I_\text{r}$ is set to 25\,µA (comparable to the valued extracted from the I-V curve) to match the experimental amplitude of $i_\text{OUT}$. We believe this double-switch behavior increased the lower limit of the margins in Figures 3b and 3c. In simulations, the effect disappears at higher frequencies (e.g. at 50\,MHz) because the input current increases fast enough to immediately generate an hotspot in both $D_3$ and $D_4$. 

% differences between diode model and experiment
The simulated output signal slightly differs from the experiment during the falling part of the pulses: in simulation, the hotspots retrap when $i_\text{IN}\approx2I_\text{r}$, while in the experiment it happens for higher input currents. This discrepancy suggests a difference in the retrapping dynamics of the hotspot between experiment and simulation. 
Indeed, the model considers the diode as a wire with asymmetric critical currents but without a notch, and thus it mimics the retrapping curve of the 400\,nm wide straight wire rather than the curve of the diode in Figure \ref{fig: 4sup}. 
As a consequence, for the model, $D_3$ and $D_4$ are quasi-ideal current sources when they are in the hotspot-retrapping regime, while in reality, they act as variable resistors and thus their current depends on the output load and voltage.  
At different frequencies and amplitudes (also in Figure \ref{fig: 14sup}), we observed that $i_\text{OUT}$ dropped when $i_\text{IN}\approx3I_\text{r}$, meaning that at that moment the input current was split almost equally between $D_3$, $D_4$, and the load. In the future, the model should be improved by reproducing the geometry to mimic the real retrapping dynamics. Moreover, Joule heating should be added to consider the experimental temperature dependence of the critical currents (see Section \ref{sec: 11sup}). 

%circulating current
The bottom panel of Figure \ref{fig: 14sup} shows also that when the input current is zero (e.g. at time zero), $i_{D1}$ and $i_{D3}$ are not zero, meaning that a circulating current remains stored in the loop after the circuit retraps in the previous period (the figure shows the waveforms at steady state). This happens due to the difference in inductance between the right and left branches of the loop. The current flows counter-clockwise at the beginning of every positive half-cycle and clockwise for every negative one.
The presence of a circulating current, the double-switch behavior, and the retrapping dynamics at steady state influence the margins for correct full-wave rectification. 
In Section \ref{sec: 15sup}, we estimated the bottom and top limits of the margins by considering all these effects and we compared the values with the experimental results.

\section{Estimation of the margins for full-wave rectification}\label{sec: 15sup}

Considering the phenomenon of double switching in Figure \ref{fig: 14sup}, and the presence of a circulating current, we can estimate the value of $|I_\text{c1}|$, lower limit of the margins for full-wave rectification: while $i_\text{IN}$ ($i_\text{IN}>0$) is increasing and a counter-clockwise circulating current $I_\text{circ}$ is already present in the loop (at steady state), $D_3$ switches at $I^+_\text{c}-I_\text{circ}$, which corresponds to $i_\text{IN}=(1+r)(I^+_\text{c}-I_\text{circ})$ with $r=L_\text{L}/L_\text{R}$; afterward, $D_3$ resets at $I_\text{r}$ and, therefore, it needs $I^+_\text{c}-I_\text{r}$ to switch again; thus, the required input current to make both $D_3$ and $D_4$ switch is $i_\text{IN} =|I_\text{c1}|= (1+r)(2|I^+_\text{c}|-I_\text{r}-I_\text{circ})$.

%top limit of the margins
The top margin depends on the negative critical current. The circuit cannot operate when the current through $D_1$ and $D_2$ is higher than $|I^-_\text{c}|$ and the input current is higher than $|I_\text{c2}|\approx|I^-_\text{c}|+I_\text{r}$, considering that a current close to $I_\text{r}$ is flowing through the other diodes.

%estimating circulating current
The experimental circulating current could not be measured, therefore, we tried to estimate it knowing $I_\text{r}$ and the input current of the circuit for which the diodes retrap ($I^\text{r}_\text{IN}$), and considering four different scenarios that may happen at every period of rectification: \\
(1) the diodes retrap simultaneously as in the simulation so $I^\text{r}_\text{IN}$ is split equally into the two branches; then the current keeps decreasing to zero, however, due to the inductance ratio $r$, most of the current is removed by the left branch so a circulating current (clockwise for $I^\text{r}_\text{IN}>0$) is generated equal to $I_\text{circ} = I^\text{r}_\text{IN}/(1+r) -I^\text{r}_\text{IN}/2= (I^\text{r}_\text{IN}/2)(1-r)/(1+r)$;\\
(2) $D_3$ resets first and current is redistributed between the two branches when $I_\text{IN}=I^\text{r}_\text{IN}$. This scenario is the most realistic because the impedance of the left branch is lower (in particular at higher frequencies), and thus more current is removed from there before the hotspot retraps. Assuming that all the current is diverted to the left branch before the right branch resets, the current in $D_3$ is $I^\text{r}_\text{IN}$ and, therefore, $I_\text{circ} = I^\text{r}_\text{IN}/(1+r)- I^\text{r}_\text{IN}= -rI^\text{r}_\text{IN}/(1+r)$;\\
(3) $D_4$ resets first and the current is diverted entirely to the right branch. the current in $D_3$ is zero and, therefore, $I_\text{circ} = I^\text{r}_\text{IN}/(1+r)$;\\
(4) the diodes reset at different times that are very close and $I^\text{r}_\text{IN}$ is redistributed unevenly between the two branches. The cause might be a combination of impedance ratio and differences in the retrapping currents of $D_3$ and $D_4$, which is plausible due to fabrication imperfections. In this scenario, $-rI^\text{r}_\text{IN}/(1+r)<I_\text{circ}<I^\text{r}_\text{IN}/(1+r)$.

% quantized current
Considering that the loop is superconducting when the diodes are not normal, the current is quantized with a step value $i_\text{step}$, which is small enough to consider the circulating current continuous: $i_\text{step}=\phi_0/L_\text{loop} \approx 400$\,nA.

% comparison with IV curve
In the measurement for Figure 3b (I-V curve of the rectifier), all four diodes were in the normal state right before the retrapping at each period (or single I-V curve measurement) and, thus no current was passing through the load at the end of the cycle. This means that $I^\text{r}_\text{IN} \approx 2I_\text{r}$. 
With this consideration and assuming that $D_3$ and $D_4$ retrapped simultaneously in the previous period, the first switching event in the upwards sweep should have happened at $|I^-_\text{c1}| = |I^+_\text{c1}| = (1+r)(2|I^+_\text{c}|-I_\text{r}-I_\text{r}(1-r)/(1+r)) = 2(1+r)|I^+_\text{c}|-2I_\text{r}$.
For $I^+_\text{c} = 104$\,µA (average between 108\,µA in Figure 2c and 100\,µA in Figure 3b with $r=1$) and $I_\text{r} = 30$\,µA (extracted from retrapping current of Figure 3b with $r=10$), we obtained 168\,µA,  in agreement with the experiment.
The second switch in the I-V curve happened around 250\,µA for both $r=1$ and $r=0.1$, suggesting that $|I^-_\text{c}| \approx 250\,\text{µA}-I_\text{r} \approx220$\,µA, a lower value than the one obtained for a single diode in Figure 2c. This discrepancy might be caused by noise in the measurement setup, not optimal field biasing in this measurement, or geometrical differences. 

%comaprison with margins vs freuency
In the margins analysis of Figure 3c the circuit could operate around 175\,µA between 100\,Hz and 10\,kHz.
This biasing condition could be associated with an unstable state achievable only at low frequencies, for which the two branches were retrapping at very close times. With $I^\text{r}_\text{IN}\approx3I_\text{r}$, the lower margin for the two diodes resetting exactly at the same time would be $|I_\text{c1}| = (1+r)(2|I^+_\text{c}|-I_\text{r}-1.5I_\text{r}(1-r)/(1+r)) \approx 155$\,µA, which is comparable to the experimental value. The presence of noise in the system contributed to an increase in the low margin, and the real circulating current could have been different from the ideal case. 

The region of operation between 218\,µA and 259\,µA at 100\,Hz and between 224\,µA and 227\,µA at 3\,MHz would be associated with the stable condition for which the left branch is always retrapping before the right branch. 
In this case, the lower limit of the margins would be $|I_\text{c1}| = (1+r)(2|I^+_\text{c}|-I_\text{r}+3rI_\text{r}/(1+r)) = 2(1+r)|I^+_\text{c}|-(1-2r)I_\text{r}\approx207$\,µA at 100\,Hz ($|I^+_\text{c}|=107$\,µA, from Figure 2c scaled by 104/108), and 239\,µA at 3\,MHz with $|I^+_\text{c}(3\,\text{MHz})|=120$\,µA (value from Figure 2c scaled by 104/108). The limits are close to the experimental results. The experimental value at 100\,Hz is higher than the estimation, probably due to noise. The experimental value at 3\,MHz is lower than the estimation, suggesting that $|I^+_\text{c}(3\,\text{MHz})|\approx115$\,µA.
The top bound of the margins correspond to $|I^-_\text{c}(100\,\text{Hz})|\approx230$\,µA and $|I^-_\text{c}(3\,\text{MHz})|\approx197$\,µA. The latter is close to 200\,µA, value from Figure 2c scaled by 104/108. The experimental top margin was lowered by noise in the system.
These results confirm that the margins shrunk with frequency due to the decrease in efficiency of the single diodes.

%margins from simulation
According to simulations, the margins at 3\,MHz with $I^+_\text{c}=115$\,µA and $I^-_\text{c}=200$\,µA were (207\,µA, 225\,µA)
introducing a circulating current $I_\text{r}(1-r)/(1+r)$ at time zero. These values are in agreement with the equations for both diodes retrapping at the same time and $I^\text{r}_\text{IN}=2I_\text{r}$. Therefore, the model overestimated the margins.
With an initial circulating current set to zero, the circuit could reach the steady state of $I_\text{circ}=I_\text{r}(1-r)/(1+r)$ and rectify only for $i_\text{IN}$ between 221\,µA and 225\,µA.

% estimating the optimal inductance ratio
With the expressions obtained above, we estimated the optimal ratio $r$ so that the current diverted after $D_3$ switches is enough to make $D_4$ switch too, and the margins are maximized (without considering the decrease of $\eta$ with frequency). The condition to have this behavior is $(1+r)(I^+_\text{c}-I_\text{circ})-I_\text{r}=I^+_\text{c}$. 
In the worst case of the left branch completely retrapping before the right one $I_\text{circ}=-rI^\text{r}_\text{IN}/(1+r)$, and therefore the condition becomes: $r=I_\text{r}/(I^+_\text{c}+I^\text{r}_\text{IN}) \approx 1/(3+I^+_\text{c}/I_\text{r})$ for $I^\text{r}_\text{IN}=3I_\text{r}$.
Using the experimental values of these currents, the result is $r_\text{opt} \approx 0.15$ ($L_\text{R}/L_\text{L} \approx 6.67$), which corresponds to a low margin of $|I_\text{c1}|_\text{opt}= (1+r_\text{opt})(I^+_\text{c}-I_\text{circ}) = I^+_\text{c}+I_\text{r}$.
However, $r$ should be slightly higher than $r_\text{opt}$ because noise in the measurement setup and fluctuations could lower the real critical current of $D_4$. 
In the future, these assumptions and calculations would need to be verified, because other dynamics we are not considering could influence the calculation of the optimal value.

\section{AC-to-DC conversion: experiment and simulation}
   \label{sec: 15}

\begin{figure*}[htbp!]
\includegraphics[width=16cm]{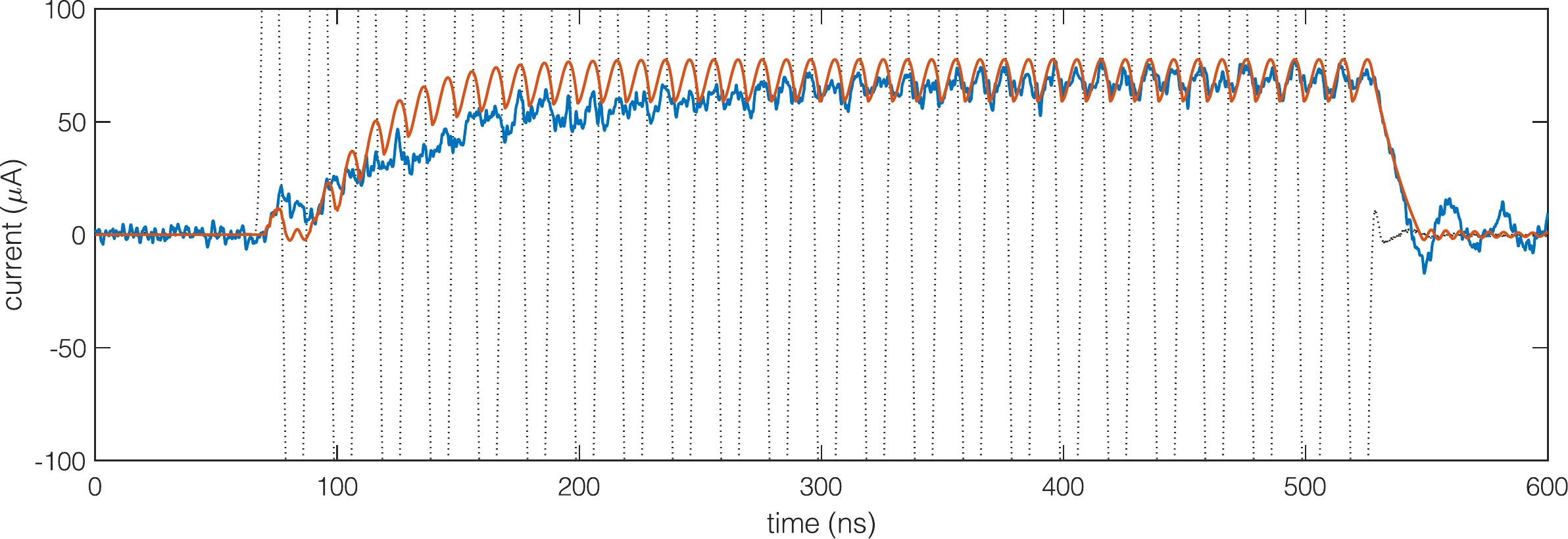}
            \caption{Comparison between experimental and simulated time-domain behavior of the AC-to-DC converter. The black dashed trace is the input current (the top and bottom portions of the signals are not shown to focus on the output trace). The blue trace is the experimental output current through the load resistor without averaging. The red trace is the simulated output current.}\label{fig: 15sup}
        \end{figure*}

Figure \ref{fig: 15sup} shows the comparison between the experimental (not averaged) and the simulated output current of the AC-to-DC converter.  
We did not use an amplifier because its limited bandwidth near DC altered the shape of the signal. 

During the first several periods of the input signal (for about 300\,ns), the rectifier did not operate correctly. 
In simulations, we observed this behavior when we took into account the inductance of the wire bonds between the PCB and the ports of the rectifier (with values in simulation on the order of 1\,nH). 
The simulated trace in this plot was obtained with zero inductances from the wire bonds. 

The simulated ripple amplitude seemed to be lower than in the experiment. Probably the reason is that we did not consider parasitic capacitances in the setup that might have altered the equivalent $C$ of the filter. For example, the capacitance of the coaxial cables on the output ports, in parallel with $C$, could have increased the $RC$ time constant of the filter.

For both the experiment and the simulation, there was a ringing transient at the falling edge of the burst. In the experiment, the ringing amplitude was higher due to the ringing behavior of the input signal.

\section{Design and simulation of bias distribution networks}

We suggest a circuit design to solve the problem of how to bias a network of superconducting devices on a chip with dynamically tunable DC currents. This design has potential value across a range of problems such as biasing detector arrays, nanocryotron electronics, and neuromorphic systems \cite{castellani_design_2020, lombo_superconducting_2022}.

The basic idea is to build integrated AC-to-DC converters to individually and dynamically set the bias level of each of an array of devices (e.g. SNSPDs). Doing this with DC cables would be challenging, however setting the DC bias levels through rectified AC signals that are frequency multiplexed on a single RF line would help decrease the number of cables coming out of the cryostat. 
Indeed, a similar frequency-multiplexed biasing and readout system has already been proposed for SNSPDs \cite{doerner_frequency-multiplexed_2017}. In this proposal, the sensors were directly biased with AC signals, thus the photon detection efficiency was time-dependent \cite{doerner_comparison_2019}. Exploiting the advantages of frequency multiplexing while maintaining a DC bias, and thus not sacrificing efficiency, would be optimal. This goal can be achieved by using integrated AC-to-DC converters to rectify the AC signals at low temperatures. The converters can be designed using superconducting bridge rectifiers.

        % why do we need bias network
        In this circuit, each cryogenic device is biased by an AC-to-DC converter. 
        Each converter is coupled to a superconducting resonator with a unique resonant frequency, and all the resonators are coupled to a single RF line coming out of the cryostat. 
        The current value of each bias line is thus frequency multiplexed. 
        With the current performance of the bridge rectifier, the maximum frequency to sustain full-wave rectification is 3\,MHz. 
        With this and lower frequency values, the footprint of the resonators would be too large. 
        Moreover, the margins would be too small to allow fine tuning.  
        Therefore, for this design, we assumed the devices had been already optimized to operate in a range of frequencies between 50\,MHz and 150\,MHz with sufficiently large margins.

\begin{figure*}[htbp!]
  
            \includegraphics[width=16.3cm]{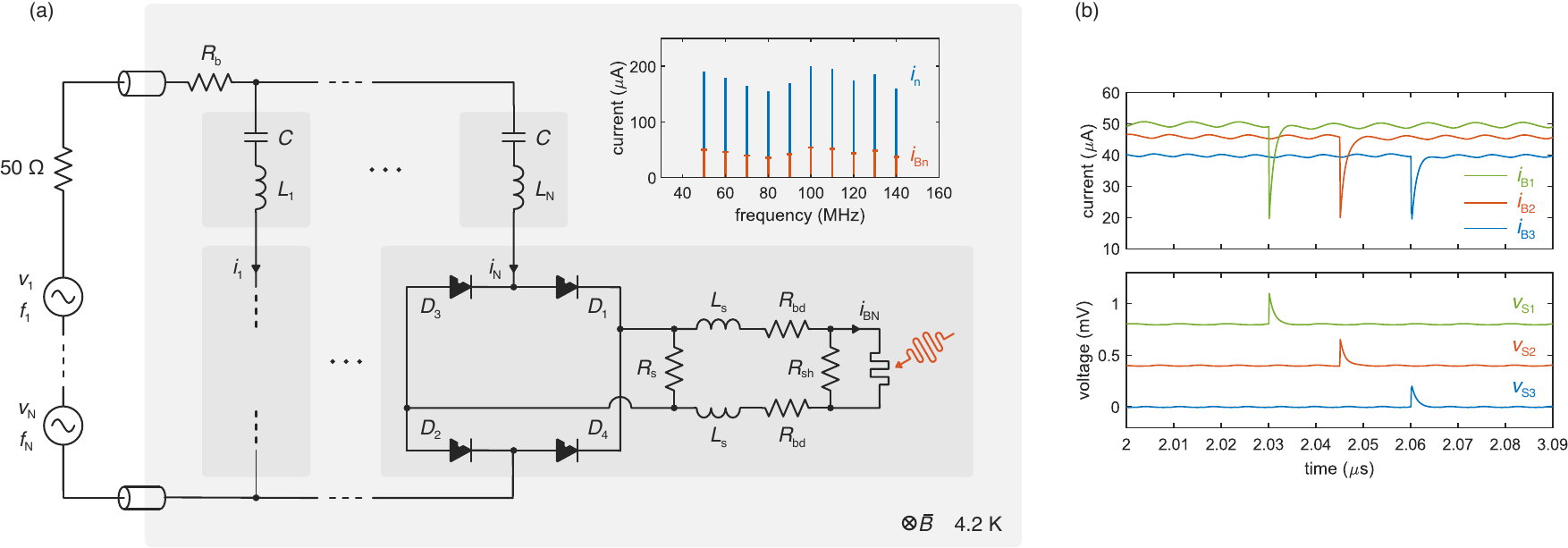}
            \caption{Design and simulation of a bias distribution network based on superconducting bridge rectifiers. (a) Circuit schematic of a network to bias $N$ SNSPDs. All the components in the gray box are placed at 4.2\,K, and the field is applied to the entire system. The voltage source at room temperature generates $N$ sinusoidal signals ($v_\text{n}$) at $N$ different frequencies ($f_\text{n}$). Each signal couples to a specific superconducting LC series resonator ($C$ and $L_\text{n}$). The resonators are connected in parallel. Each resonator is in series with a bridge rectifier, which drives an inductive low-pass filter. The filter generates a DC current passing through a shunted SNSPD. The upper-right inset shows the distribution of different bias levels for $N=10$. The amplitude of the AC input current $i_\text{n}$ of each rectifier is in blue. The DC current through the SNSPD, with associated ripple (error bars), is in red. (b) Simulated time-domain behavior for three SNSPDs (respectively coupled to signals at 50\,MHz, 60\,MHz, 70\,MHz) switching at different times in the network of ten SNSPDs. Upper panel: current through the SNSPDs. lower panel: voltage $v_\text{Sn}$ across the shunt resistor $R_\text{sh}$ of the SNSPDs. Traces in the lower panel are vertically shifted for clarity. Circuit parameters: $I^{+}_\text{c} = 100$\,µA, $I^{-}_\text{c} = 200$\,µA, $R_\text{b}=45$\,$\Omega$, $C = 1$\,pF, $R_\text{s}=5$\,$\Omega$, $L_\text{s}=250$\,nH, $R_\text{sh}=10$\,$\Omega$, $R_\text{bd}=2.5$\,$\Omega$, SNSPD inductance: $L_\text{d}=10$\,nH.}\label{fig: 16sup} 
        \end{figure*}

        % architecture
        Figure \ref{fig: 16sup}a shows the circuit schematic of the proposed architecture for SNSPD biasing. 
        A single RF line is used to send $N$ frequency components to the associated $N$ series lumped LC resonators, which are connected in parallel to each other (LC circuits might be replaced by CPW resonators as in the work of Doerner et al. \cite{doerner_frequency-multiplexed_2017}).
        The capacitance $C$ is constant for all the resonators, $L_\text{n}$ varies.
        In series with each resonator, there is an AC-to-DC converter that biases a single SNSPD.
        For each frequency component of the input signal, only one branch is at resonance, and thus ideally, the AC signal is delivered to only one converter. 
        The converter is composed of a bridge rectifier and an inductive low-pass filter made with $L_\text{s}$, $R_\text{s}$ and $R_\text{L}$. We opted for this solution because superconducting nanowire-based kinetic inductances have much lower footprints than integrated capacitors. The SNSPD is in series with $R_\text{L}$ and shunted by $R_\text{sh}$. 
        The system is fully differential and the values of $R_\text{b}$ and $R_\text{s}$ are chosen so that their sum is 50\,$\Omega$ to ensure impedance matching with the room-temperature voltage generator at each frequency.   

        %readout for snspds
        For this architecture, we chose to bias SNSPDs as example devices but the scheme can be used for other superconducting devices. 
        It is worth using this structure for SNSPDs only if additional superconducting circuitry is added to read out the detector's outputs and eventually frequency multiplex them, such that there is only one readout line.
        A possible solution might be to use a single RF line for both frequency-multiplexed biasing and readout, similar to Doerner et al. \cite{doerner_frequency-multiplexed_2017}
        To do so the low-pass filter should be redesigned such that the impedance seen by the input voltage source drastically increases after a detection event and the impedance change can be measured on the input port. 
        Another readout solution can be to thermally couple the detectors or their shunt resistors to a common superconducting line, similar to the configuration of Oripov at al.\cite{oripov_superconducting_2023} 
        Each section of the common line coupled to a different detector could be shunted with a different value of resistance so that the output would be amplitude-multiplexed, as proposed by Gaggero et al. \cite{gaggero_amplitude-multiplexed_2019}
        As a third solution, SNSPDs could directly drive nanocryotron electronics that perform signal processing at low temperatures, removing the need to acquire the detector's output with room-temperature equipment. 

        %simulations
        We simulated the system of Figure \ref{fig: 16sup}, without considering readout architectures, in LTspice by using the model of the rectifier with a constant 33\,\% rectification efficiency, $L_\text{L} = 200$\,pH, and $L_\text{R} = 2$\,nH.
        The inset in Figure \ref{fig: 16sup}a shows the different levels of input current of the bridge rectifier set by the $N$ components of the voltage source (blue bars), and the associated DC output currents (red bars), for $N=10$. 
        The values of input currents were chosen arbitrarily inside the input margins.
        The DC currents follow the distribution profile of the input signals because the bias levels of the 10 devices can be individually controlled. 
        The range of possible bias levels goes from 35.5\,µA to 54\,µA.

        % simulation of 3 snspds
        Figure \ref{fig: 16sup}b shows the simulation result of 3 out of the 10 SNSPDs detecting photons at different times. 
        The detectors are biased with three different values of current that are set by the voltage sources. 
        After the detection events, the bias current is reset to the initial value and the detectors are ready to fire again. 
        The switching event of one of the detectors does not alter the operation of the other devices. 

        %start-up time
        The system reaches a steady state with stable DC currents after about 1\,µs (start-up time) from when the input voltage source is turned on. 
        After this time, the value of one of the bias currents can be updated from the highest to the lowest level of the margins in about 200\,ns. 
        During the update the current alteration of the other channel is negligible. 

        %deltaf
        In the simulated circuit, the 10 channels are separated in frequency by $\Delta f =10$\,MHz. 
        To scale up the system (increasing $N$), $\Delta f$ will need to be lowered if the range of allowed frequencies is fixed (assuming that the rectifier operates only between 50\,MHz and 150\,MHz with reasonably small footprints). 
        However, decreasing $\Delta f$ increases the cross-talk between resonators and thus the effective ripple of the DC currents. 
        If the ripple amplitude becomes comparable with the current margins of the rectifier, fine-tuning the currents is no longer possible. 
        With the parameters used, this condition is reached for $\Delta f < 6.5$\,MHz. 
        With $\Delta f = 6.5$\,MHz the largest ripple is about 15\,\% of the total margin. 
        With $\Delta f = 10$\,MHz the larger ripple is about 9.5\,\% of the total margin. 
        Considering these results, and considering that the rectifier can operate up to 150\,MHz, the current design will not allow more than 15 channels.
        
        % power consumption
        We estimated the power consumption of this system in simulations. The average consumption at low temperatures is 810\,nW per device, without switching events of the biased devices.
        92\,\% of this power is dissipated by $R_\text{b}$, 7\,\% of the power is dissipated by the bridge rectifier and the filter, and 1\,\% is dissipated by the $R_\text{bd}$ bias resistors. 

        %footprint
        The designed architecture would have a relatively large footprint when implemented. 
        The area of each module would be dominated by the capacitor $C$ and the inductor $L_\text{n}$. 
        If rectifiers operating at higher frequencies were demonstrated, the footprint of the resonators could be decreased. 
        Otherwise, a multi-layer fabrication process might be used to minimize the occupied area. 
        In particular, parallel-plate capacitors could be used.
        Moreover, a normal metal layer should be included for the integrated resitors. 
        
        Another circuit element that largely contributes to the occupied area is the large inductor in the low-pass filter. 
        Its inductance can be in principle reduced, however, the ripple on the output current would increase.

\providecommand{\latin}[1]{#1}
\makeatletter
\providecommand{\doi}
  {\begingroup\let\do\@makeother\dospecials
  \catcode`\{=1 \catcode`\}=2 \doi@aux}
\providecommand{\doi@aux}[1]{\endgroup\texttt{#1}}
\makeatother
\providecommand*\mcitethebibliography{\thebibliography}
\csname @ifundefined\endcsname{endmcitethebibliography}  {\let\endmcitethebibliography\endthebibliography}{}